\documentstyle[twoside,fleqn,espcrc2,psfig]{article}

\newcommand{\AmS}{{\protect\the\textfont2
  A\kern-.1667em\lower.5ex\hbox{M}\kern-.125emS}}

\newcommand{\be}{\begin{equation}}
\newcommand{\ee}{\end{equation}}
\newcommand{\bea}{\begin{eqnarray}}
\newcommand{\eea}{\end{eqnarray}}

\def\ltap{\;\raisebox{-.5ex}{\rlap{$\sim$}} \raisebox{.5ex}{$<$}\;}
\def\gtap{\;\raisebox{-.5ex}{\rlap{$\sim$}} \raisebox{.5ex}{$>$}\;}

\newcommand{\ia}{a^{-1}}
\newcommand{\fb}{f_B}
\newcommand{\fbs}{f_B^{stat}}
\newcommand{\zren}{Z^{Ren}}
\newcommand{\gap}{$\!\!\!\!\!\!\!\!\!\!$}

\title{Leptonic Decays of Heavy-Light Systems}

\author{Chris Allton\address{Dipartimento di Fisica,
                             Universit\`a di Roma ``La Sapienza'',
                             and INFN, Sezione di Roma I, \\
                             P.le Aldo Moro,
                             00185 Roma.}
        \thanks{Address from 1st October, 1995:
                Department of Physics,
                University of Wales, Swansea,
                Singleton Park,
                Swansea SA2 8PP,
                U.K.
                Preprint Rome 95/1114 \& SWAT/90}}
       
\begin{document}

\begin{abstract}
Results from recent lattice calculations of the decay constants
$f_B$ and $f_D$ are reviewed.
A discussion of the methods currently used is presented,
together with an outline of the various systematic effects involved.
\end{abstract}

\maketitle

\section{Introduction}

Weak matrix elements of heavy-light mesons are of fundamental
importance in particle physics since they enter determinations
of some of the least known CKM matrix elements, and of the
$B-\bar{B}$ and $D-\bar{D}$ mass splittings.
The simplest matrix element to study is the leptonic decay
constant of a pseudoscalar meson, $f_P$. It is defined:

\[
<0|A_\mu|P> = f_P p_\mu,
\]
where $|P>$ is a pseudoscalar meson with 4-momentum $p_\mu$,
and $A_\mu$ is the axial current.

The feasibility of using the lattice technique to calculate $f_P$
is now firmly established.
The general trend in lattice calculations of $f_P$ is now towards
a greater understanding of the systematic effects entering the
calculation. This has been made possible through smaller
statistical errors uncovering systematics which were previously
hidden. Thus, the effects of, e.g., quenching, different
choices of interpolating operators etc., can be studied.
Systematic effects will be a focus throughout much of this review.

The plan of this review is as follows. The next section overviews
the three methods currently employed by the lattice community
to calculate $f_P$. Recent results from each of these methods
are then presented in the subsequent sections.
The present status of lattice calculations is summarised
in the conclusion.

Unless explicitly stated, all results are to be taken as
``preliminary''.

\section{Overview of Methods Used}

Heavy Quark Effective Theories (HQET) (see e.g. \cite{hqet})
are invaluable tools in the study of the spectrum and decays of systems
involving one or more heavy quarks.
In this theory, the QCD action is systematically expanded
in terms of the inverse heavy quark mass, $m_Q$.

HQET is extremely powerful in lattice calculations
as well as in the continuum.
It provides a means of overcoming the problem of simulating
quarks with mass greater than the inverse lattice spacing.
In general, a lattice
calculation using HQET proceeds much as in conventional
lattice calculations, but with all heavy quark propagators
calculated using HQET.
Thus a heavy-light pseudoscalar meson correlation function
in the conventional approach would be, e.g.
\bea\nonumber
C(t)&=&\sum_{\vec{x}} <A_0(x) A_0(0)> \\\nonumber
      &=&\sum_{\vec{x}} Tr \{ S_q(x,0) \gamma_5 S_Q(0,x) \gamma_5 \}
\eea
where $S_q$ is the light quark propagator
calculated using the traditional discretised Dirac equation,
and the heavy quark propagator, $S_Q$, is calculated using the discretised
HQET action up to a chosen order in $1/m_Q$.

In this review I will be discussing results obtained with $S_Q$
calculated using

\begin{itemize}
\item the zeroth order HQET (also termed the `static' case)
\item HQET to some finite (typically the first) order (also termed NRQCD).
\end{itemize}
I will also discuss results with $S_Q$ determined from
\begin{itemize}
\item the conventional lattice Dirac equation (which I will term the
``conventional'' approach).
\end{itemize}

The next three sections discuss recent results from each of
the above approaches in turn.
The lattice actions or $S_Q$'s used in each case are defined
in the corresponding sections.

\begin{figure}
\protect\label{fig:frm}

\begin{picture}(220,170)(-20,0)

\put(0,0){\vector(1,0){165}}
\put(170,-2){$1/M_P$}
\put(0,0){\vector(0,1){150}}
\put(-20,160){$f_P\sqrt{M_P}$}

\put(20,-4){\line(0,1){8}}
\put(18,-15){0}
\put(50,-4){\line(0,1){8}}
\put(40,-15){$1/M_B$}
\put(100,-4){\line(0,1){8}}
\put(90,-15){$1/M_D$}

\put(20,90){\line(0,1){40}}
\put(8,135){Static}

\put(20,110){\line(1,0){25}}
\multiput(20,110)(5,0){12}{\line(1,0){2.5}}
\put(40,115){NRQCD}

\put(160,80){\line(-1,0){80}}
\multiput(160,80)(-5,0){22}{\line(-1,0){2.5}}
\put(55,85){Improved-Conventional}

\put(160,45){\line(-1,0){30}}
\multiput(160,45)(-5,0){12}{\line(-1,0){2.5}}
\put(105,60){Unimproved-}
\put(105,50){Conventional}

\end{picture}


\caption{Representation of the range of validity
of the three methods discussed in this review.}
\end{figure}
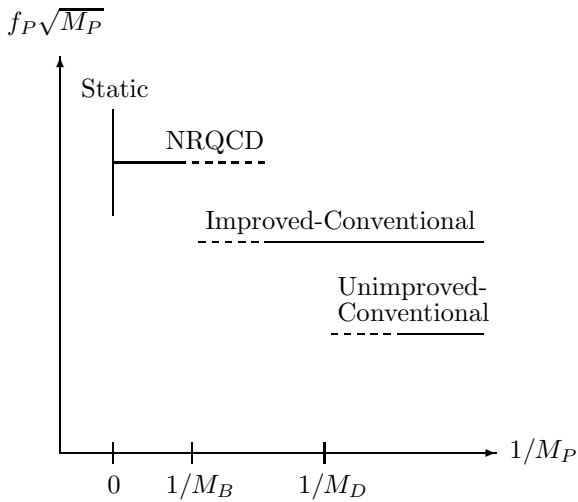


Some general comments regarding the applicability of the
three approaches are helpful at this stage.
One can represent the range of validity of each of the
above methods graphically in fig.1. 
Here $f_P \sqrt{M_P}$ is plotted against $M_P$.
Note that in HQET it can be shown that, up to unimportant logarithms,

\[
f_P \sqrt{M_P} = constant + O(1/M_P).
\]
The static method is valid only for $m_Q = \infty$
and serves as a reference point for extrapolations in $1/m_Q$.
The NRQCD approach \cite{nrqcd,beth} is valid for $1/(m_Q a) << 1$,
and, at $O(1/m_Q)$ its range of validity
should include masses down to around $M_B$.
The conventional approach, for present lattice parameter
values, is believed to be accurate for meson masses
of around $M_D$ and less.
In fact, this statement assumes that some improvement
scheme such as the SW action \cite{sw} (hereafter termed
the ``Clover'' action) and/or the ``Fermilab'' (also termed
the ``Heavy'') formalism \cite{heavy,paul,andreas,tobe}
is implemented.
If no such improvement is undertaken, then the conventional
approach fails for $M_P \gtap M_D$.

Taken together, these three approaches span the entire range
of $m_Q$, from infinity to the charm quark mass, $m_c$ and below.
Clearly, a necessary condition for the success of lattice calculations
is that these three methods agree in the shared regions, and
that they provide a continuous functional behaviour for the
quantities measured.
It is one of the purposes of this review to study the status of this
consistency check.

To set the notation, lattice calculations of $f_P$ determine the
dimensionless, unrenormalised quantity $\phi^\#$ corresponding to $f_P$.
It is defined (using an obvious notation):

\be
f_P \sqrt{M_P} = Z^{Ren} \phi^{\#} a^{-3/2}.
\label{eq:frm_lat}
\ee
(Note other notation for $\phi^\#$ appearing in the literature
includes $\sqrt{2} Z_L$ and $\sqrt{2} \tilde{f_B}$.)
I have chosen the superscript $^\#$ to denote unrenormalised,
dimensionless lattice quantities.


\section{Static Results}
\label{stat}

The static action has a very simple form
and allows the heavy quark to propagate only in time,
and not in position space.
This action can be inverted to give the following form
for $S_Q$ \cite{eichten,ef}:

\[
S_Q(x,0) = {\cal P}_{\vec{x}}(t_x) \; \; \delta(\vec{x}) \;
\theta(t_x)\; e^{-m_Q t_x}\;
\frac{1+\gamma_4}{2}
\]
where ${\cal P}_{\vec{x}}(t_x)$ is the Polyakov line from the
point $(\vec{x},0)$ to $(\vec{x},t_x)$.

\subsection{Smearing}
\label{smear}

The history of the static approach on the lattice has been
plagued by problems associated with its poor signal to noise ratio.
The reason for this effect is now well understood
in terms of the variance of the correlation function
picture \cite{variance}.
An approach to circumvent this problem is to introduce
smeared interpolating operators \cite{ph}.
These can be expressed as

\[
A_0^S(\vec{x},t) = \sum_{\vec{y},\vec{z}} \overline{Q}(\vec{y},t)
  \gamma_0 \gamma_5 q(\vec{z},t)
                                    \psi(|\vec{x}-\vec{y}|,|\vec{x}-\vec{z}|).
\]
$A_0^S$ still couples to the ground state because it has the same
quantum numbers as the local current $A_0^L$, but using it to define
correlation functions can result in a smaller overlap with the excited states.

There is a wide variety of forms of the smearing function $\psi$ presently
in use.
The most simple form of smearing functions is `cubic' defined
\cite{elc_60}:
\bea\nonumber
\psi(\vec{y},\vec{z}) &=& \delta(\vec{z}) ; \\\nonumber
                      & & \mbox{for all $|y_i| \le L_s/2$ ;
                             i=1,2,3} \\
                      &=& 0 \;\; \mbox{otherwise.}
\label{eq:cube}
\eea
These have the advantage that they are easiest to code, especially
on parallel machines, but have the disadvantage that
they require a gauge fixing procedure and are
a rather brutal approximation to the physical wave function.
This second point will be discussed more in sec \ref{ape}.

Another popular choice of smearing functions is based on the
non-relativistic quark model \cite{fnal}. Here, the heavy quark
potential of the gauge configurations in the simulation
is used to obtain the quantum mechanical wave function.
Its advantage is that the physics is being used to dictate the
smearing.

The Wuppertal collaboration have proposed various smearings
\cite{lat89_stefan,wup}.
The advantage in these cases is that the smearing is relatively
quick, and is gauge invariant.

My favourite smearing is ``MOST'' (Maximal Operator Smearing Technique)
\cite{ken1}. In this technique the set of operators is defined where
the $Q$ and $q$ fields have {\em every} possible separation
(which is independent under cubic symmetry).
The two-point correlation between all members of this set is measured,
and an simple analysis \cite{kron,lw} can be used to extract
the ground state properties.
The advantage of this method is obvious: because it uses all possible
relative separations as a basis, all other smearing
methods are a subset of this approach.
Its use in light-light physics should therefore be investigated.
The disadvantage is that it becomes memory intensive for
large lattices.

Note that in order to extract the local matrix element, $f_P$,
local-smeared, as well as smeared-smeared correlation functions
need to be measured. In practice it is best
to place the local current at the sink rather than the source, since a
better signal to noise ratio results \cite{lat89_eichten,wup}.

\subsection{\mbox{\boldmath$Z^{Ren}$}}
\label{zren_stat}

In order to define the physical $f_P$ a choice needs to be
made for the definition of the renormalisation constant $Z^{Ren}$
see eq.(\ref{eq:frm_lat}).
Perturbative calculations \cite{zastat_wil} show,
in the case of the static-Wilson current,

\be
\zren_{stat}(g^2) = (1 - .189 g^2) (1 + .0127 g^2 log(a^2 m_Q^2)).
\label{eq:zren_stat}
\ee
For the static-Clover case, $0.189$ is replaced by $0.144$
\cite{zastat_clo}. Note that the first factor
is essentially the lattice to continuum
matching within the effective (static) theory, and the second is the
matching between the effective and full theory in the continuum.

Using the Kronfeld-Lepage-Mackenzie formalism \cite{lm,andreas}
leads to the following
definition of $\zren_{stat}$ for the Wilson action,
\bea\nonumber
\zren_{stat}(g^2) &=&
 \frac{\sqrt{1-6\tilde{\kappa_q}}}{\sqrt{2\kappa_q}}
                       \times (1 - 0.137 \tilde{g}^2) \\
                  & & \times  (1 + 0.0127 \tilde{g}^2 log(a^2 m_Q^2)).
\label{eq:zren_stat_lm}
\eea
Here, the renormalised couplings are $\tilde{g}^2 = 6/\beta / u_0^4$
and  $\tilde{\kappa_q} = u_0 \kappa_q$. $u_0$ will be defined later.
The first factor in eq.(\ref{eq:zren_stat_lm})
is the rescaling of the light quark field from the traditional,
Wilson $\sqrt{2\kappa_q}$ to the KLM prescription.
There is however an ambiguity in the above definition of $\zren_{stat}$.
This is because one is free to define $u_0$ as, e.g. the average plaquette,
$<U_{Plaq}>^{1/4}$, or $1/8\kappa_c$ (where $\kappa_c$ is the critical value of
the hopping parameter).

Table \ref{tab:zren_stat} displays the values of $\zren_{stat}$ obtained
with the above definitions of $\zren_{stat}$.
As can be seen from the table the difference between the various
determinations of $\zren$ is $O(30\%)$ and it is $O(5\%)$ even
within the KLM formalism.
Obviously, this means that the overall systematic error in $\fbs$ has a
contribution of this order from the uncertainties in $\zren$.
This error from the determination of $\zren_{stat}$ has now become
one of the dominant errors in the lattice calculations of $\fbs$.
The way around this problem is to use 
a non-perturbative definition of $\zren$ which are becoming
available which circumvent the above problems completely
\cite{lat95_mauro}.


\begin{table}
\caption{$\zren_{stat}$ obtained using different definitions.
Column 2 uses the ``naive'' definition of $g^2 = 6/\beta$
(see eq.(\ref{eq:zren_stat})).
Columns 3 and 4 use the KLM formulation
(at the chiral limit) with the $u_0$ definitions given by
$1/(8\kappa_c)$ and $<U_{plaq}>^{1/4}$ respectively
(see eq.(\ref{eq:zren_stat_lm})).}
\label{tab:zren_stat}
\begin{center}
\begin{tabular}{lrrr}
\hline
$\beta$             & \multicolumn{3}{c}{$\zren_{stat}$} \\
                      \cline{2-4}
\hline
6.0                 & 0.83  & 0.62   & 0.59   \\
6.1                 & 0.83  & 0.64   & 0.61   \\
6.2                 & 0.83  & 0.66   & 0.63   \\
6.3                 & 0.83  & 0.67   & 0.64   \\
6.4                 & 0.83  & 0.68   & 0.65   \\
\hline
\end{tabular}
\end{center}
\end{table}

\subsection{Continuum Limit}
\label{cont_limit}

The possible variation of $f_P$ in the static limit
with the lattice spacing, $a$, has been long discussed
(see, for example, \cite{lat93_claude,wup,fnal}
This so-called {\em non-scaling} behaviour has been
proposed as a mechanism for reducing the lattice
prediction for $f_P$ by as much as 30\%.
In principle this $a$ dependence is simple to study.
One simply calculates $f_P$ on the lattice using a
number of $\beta$ values and extrapolates the results
to $a=0$. In practice this method is difficult to implement
due to the increasing statistical errors of the data points
closest to the point of interest, $a=0$.

Recently a method has been proposed to analyse the scaling
behaviour of $f_P$, and indeed any other dimensionful
lattice quantity \cite{cra}.
Instead of studying the scaling of the final quantity,
$f_P$, the scaling of each of the three
factors in eq.(\ref{eq:frm_lat}) which define $f_P$:
$\zren_{stat}$; $\phi^\#$; and $\ia$ are determined separately.
Each of these three factors are functions of $g^2$.
The question to be answered is: does the $g^2$
dependence of these factors cancel in the overall product?
The analysis performed in \cite{cra} suggests that,
within present statistical errors, $f_P$ {\em does scale}
but only for $\beta \gtap 6.0$.
This means that analyses which use a linear fit in $a$
and data with $\beta \ltap 6.0$ are biasing their continuum
extrapolation to a smaller value.
This is due to the levering effect
of the non-scaling, and typically higher data for $\beta \ltap 6.0$,
which, furthermore, typically have smaller statistical errors.

The analysis in \cite{cra} also showed that the dominant
non-scaling effect was not due to higher orders in perturbation
theory in $g^2 = 6/\beta$, but rather to $O(a)$ effects.
A common $20\%$ to $30\%$ discretisation effect was found
for the quantities studied, such as $\phi^\#$, $M_\rho$ and $f_\pi$.

\begin{table*}[hbt]
\setlength{\tabcolsep}{1.5pc}
\newlength{\digitwidth} \settowidth{\digitwidth}{\rm 0}
\catcode`?=\active \def?{\kern\digitwidth}
\caption{Summary of recent results using the static approach.
`Q' signifies Quenched simulation, `D' Dynamical.
See text for detailed comments.
All numbers should be considered preliminary with the exception
of [30].} 
\label{tab:stat}
\begin{tabular*}{\textwidth}{@{}l@{\extracolsep{\fill}}cccc}
\hline
{\bf Collaboration} & UKQCD & SGO    & MILC   & APE \\
Reference           &\cite{lat95_dgr,ukqcd}
                            & \cite{lat95_sara,lat95_arifa}
                                     & \cite{lat95_claude}
                                              & \cite{ape_64} \& see text \\
\hline
\multicolumn{5}{l}{{\bf $\!\!\!\!\!\!\!\!\!\!$Lattice Parameters}}  \\
$n_F$               & 0	    & 0 \& 2 & 0 \& 2 & 0 \\
$\beta$             & $6.2$ & $5.6$ D& $5.7 \rightarrow 6.5$ Q & $6.0,6.1,6.2,6.4$ \\
                    &       & $6.0$ Q& $5.4 \rightarrow 5.7$ D & \\
Volume              & $24^3 \times 48$
                            &        & $\le 32^3 \times 100$
                                              & $\le 24^3 \times 64$ \\
$N_{cfgs}$          & $60$  & $100$ D& $O(100)$& $\ltap 400$ \\
                    &       & $35$ Q &        & \\
$S_q$               & Clover& Wilson \& Clover& Wilson & Wilson \& Clover \\
$S_Q$               & Static& Static & Static & Static \\
                    &       &\& NRQCD&\& Conventional& \\
\hline
{\bf Results}       &  &        &        & \\
$\fbs$ [MeV]        & $266^{+18+28}_{-20-27}$
                            & \multicolumn{1}{c}{see Section \ref{nrqcd}}
                            & \multicolumn{1}{c}{see Section \ref{conv}}
                                              & $240(30)$ \\
$f_{B_S}^{stat}/\fbs$ & $1.16^{+4}_{-3}$
                            & \multicolumn{1}{c}{"}
                                      & \multicolumn{1}{c}{"}
                                              & $1.17(4)$ \\
$M_{B_S}-M_B$ [MeV] & $87^{+15+6}_{-12-12}$
                            & \multicolumn{1}{c}{"}
                            & \multicolumn{1}{c}{"}
                                              & $81(10)$ \\
\hline
\end{tabular*}
\end{table*}

\subsection{Recent Results}

Recent results for $f_B^{stat}$ are shown in table \ref{tab:stat}
together with their references in the second row.
A striking fact about these latest simulations is the
appearance of simulations with dynamical fermions
\cite{lat95_sara,lat95_claude}.
A discussion of the effects of this on $f_P$ appears in
secs. \ref{sgo_dyn},\ref{milc} and in the conclusion.

A brief description of each of the groups' work,
together with a longer discussion on the
results from the APE collaboration follows.
I have chosen to present a relatively detailed
discussion of the APE collaboration's work
since it will not be presented elsewhere in the proceedings.

The UKQCD Collaboration \cite{lat95_dgr} presented results
for both $f_B^{stat}$ and $B_B^{stat}$ (see also \cite{ukqcd}).
They noted that the systematic error associated with $\zren$ is now
a major uncertainty (see sec.\ref{zren_stat}).
A summary of their results is shown in table \ref{tab:stat}.
(The first error is statistical and the second systematic.)

The SGO collaboration \cite{lat95_arifa,lat95_sara} used
both the static and NRQCD implementation of the heavy quark
propagator, and their results will be discussed in detail
in the NRQCD section \ref{nrqcd}.

The MILC collaboration \cite{lat95_claude} combine an analysis
of both the static and conventional approaches, and a
discussion of their work will appear in the latter section
(sec. \ref{milc}).

\subsection{APE Results}
\label{ape}

The APE collaboration has recently carried out a high
statistics calculation of $\fbs$ in the quenched
approximation \cite{ape_64}.
This work is a continuation of a long program of work
in this area \cite{ape_6018,ape_sme0,lat94_cra}.
As discussed in sec. \ref{smear}, the cubic smearing
function was used. The methods used in the past for
extracting the local matrix element $\phi^{\#}$ from
the smeared correlation functions are discussed
in \cite{ape_sme0}.
One problem with these methods is that they require the
selection of the ``best'' cube size $L_s$ (see eq.(\ref{eq:cube})).
A number of techniques were developed to make this choice
as free from bias as possible \cite{ape_sme0,lat94_cra}.
However, the problem remained that cubes are available at
only discrete sizes, and that as the light quark
mass was varied, the meson's variation in size could not be adequately
tracked by the cubes.
For example, at $\beta=6.4$ the best cube size for the heaviest
of the light quarks studied was clearly $L_s=15$, but for the lightest
quark was somewhere between $L_s=15$ and $L_s=17$.

A solution to this problem has already been proposed by a number
of authors (see e.g. \cite{fnal}):
one includes various cube sizes
altogether in the one $\chi^2$ fit so that the fit is not
constrained by a single cube size.
Thus, in the example considered above for $\beta=6.4$,
the new enlarged $\chi^2$ is

\[
\chi^2 = \!\!\sum_{L_S=15,17} \sum_t \sum_{i=SS,LS}
 \left(\frac{C^{L_S,i}(t) - f^{L_S,i}(t)}{\sigma^{L_S,i}(t)}\right)^2.
\]
In this definition, both the $L_s=15$ and $L_s=17$ cubes
are included in the $\chi^2$ as well as the usual
sum over the times $t$ in the fitting window.
The sum $i=SS,LS$ includes the contributions from the
smeared-smeared and the smeared(at source)-local(at sink) correlation
functions.
The fitting function $f^{L_S,i}(t)$ is chosen appropriately;
for example,
\bea
f^{L_S=15,i=LS}(t) &\equiv& \frac{\phi^{\#} \phi^\#(15)}{2} e^{-Mt}
\times \nonumber\\
                   &      & ( 1 + \eta(L) \eta(15) e^{-\Delta t} ) 
\eea
where a two state fit is performed and the $\eta$'s and
$\Delta$ are the overlap and mass gap of the first excited state
respectively.

The APE Collaboration has re-done its analysis for its whole
data set at $\beta=6.0,6.1,6.2$ and $6.4$ using the above
method. As an example, fig.2 
shows the results of the new method together with the previous
analysis in the case of $\beta=6.4$.
As can be seen, the use of the new method results in an
increased slope in the chiral extrapolation.
Thus the chiral ratios and differences, $f_{B_S}^{stat}/\fbs$
and $M_{B_S}-M_B$ will increase.
The overall effect is not large,
but, {\em because the statistical errors are getting smaller},
systematic effects such as this are becoming important.

In fig.3, 
the values for $\fbs$ at various
$\beta$ values are plotted against the lattice spacing $a$
using the new method.
In this plot, the rho mass was used to set the scale.
A preliminary fit to a constant (see sec.\ref{cont_limit})
results in $\fbs = 240(30)$ MeV.
Values of $f_{B_S}^{stat} / \fbs = 1.17(4)$, and
$M_{B_S} - M_B = 81(10)$ MeV (compared to $96(6)$ Mev from experiment)
are also predicted.
All errors quoted here include both the statistical
and $a\rightarrow 0$ uncertainties.
Finite volume effects are believed to be negligible for these
lattices \cite{wup}.

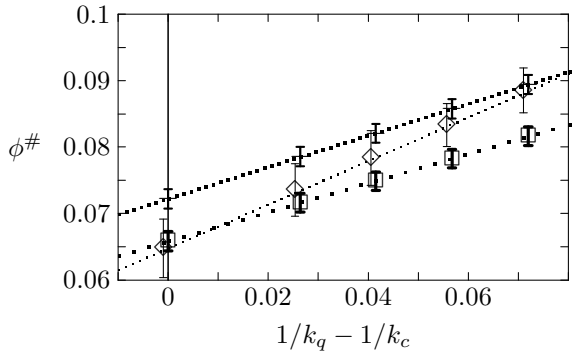
\begin{figure}
\protect\label{fig:phi_ape}

\setlength{\unitlength}{0.14pt}
\ifx\plotpoint\undefined\newsavebox{\plotpoint}\fi
\sbox{\plotpoint}{\rule[-0.200pt]{0.400pt}{0.400pt}}%
\special{em:linewidth 0.4pt}%
\begin{picture}(1500,900)(-80,0)
\font\gnuplot=cmr10 at 10pt
\gnuplot
\put(355,113){\special{em:moveto}}
\put(355,832){\special{em:lineto}}
\put(220,113){\special{em:moveto}}
\put(240,113){\special{em:lineto}}
\put(1436,113){\special{em:moveto}}
\put(1416,113){\special{em:lineto}}
\put(198,113){\makebox(0,0)[r]{0.06}}
\put(220,203){\special{em:moveto}}
\put(240,203){\special{em:lineto}}
\put(1436,203){\special{em:moveto}}
\put(1416,203){\special{em:lineto}}
\put(220,293){\special{em:moveto}}
\put(240,293){\special{em:lineto}}
\put(1436,293){\special{em:moveto}}
\put(1416,293){\special{em:lineto}}
\put(198,293){\makebox(0,0)[r]{0.07}}
\put(220,383){\special{em:moveto}}
\put(240,383){\special{em:lineto}}
\put(1436,383){\special{em:moveto}}
\put(1416,383){\special{em:lineto}}
\put(220,473){\special{em:moveto}}
\put(240,473){\special{em:lineto}}
\put(1436,473){\special{em:moveto}}
\put(1416,473){\special{em:lineto}}
\put(198,473){\makebox(0,0)[r]{0.08}}
\put(220,562){\special{em:moveto}}
\put(240,562){\special{em:lineto}}
\put(1436,562){\special{em:moveto}}
\put(1416,562){\special{em:lineto}}
\put(220,652){\special{em:moveto}}
\put(240,652){\special{em:lineto}}
\put(1436,652){\special{em:moveto}}
\put(1416,652){\special{em:lineto}}
\put(198,652){\makebox(0,0)[r]{0.09}}
\put(220,742){\special{em:moveto}}
\put(240,742){\special{em:lineto}}
\put(1436,742){\special{em:moveto}}
\put(1416,742){\special{em:lineto}}
\put(220,832){\special{em:moveto}}
\put(240,832){\special{em:lineto}}
\put(1436,832){\special{em:moveto}}
\put(1416,832){\special{em:lineto}}
\put(198,832){\makebox(0,0)[r]{0.1}}
\put(220,113){\special{em:moveto}}
\put(220,133){\special{em:lineto}}
\put(220,832){\special{em:moveto}}
\put(220,812){\special{em:lineto}}
\put(355,113){\special{em:moveto}}
\put(355,133){\special{em:lineto}}
\put(355,832){\special{em:moveto}}
\put(355,812){\special{em:lineto}}
\put(355,68){\makebox(0,0){0}}
\put(490,113){\special{em:moveto}}
\put(490,133){\special{em:lineto}}
\put(490,832){\special{em:moveto}}
\put(490,812){\special{em:lineto}}
\put(625,113){\special{em:moveto}}
\put(625,133){\special{em:lineto}}
\put(625,832){\special{em:moveto}}
\put(625,812){\special{em:lineto}}
\put(625,68){\makebox(0,0){0.02}}
\put(760,113){\special{em:moveto}}
\put(760,133){\special{em:lineto}}
\put(760,832){\special{em:moveto}}
\put(760,812){\special{em:lineto}}
\put(896,113){\special{em:moveto}}
\put(896,133){\special{em:lineto}}
\put(896,832){\special{em:moveto}}
\put(896,812){\special{em:lineto}}
\put(896,68){\makebox(0,0){0.04}}
\put(1031,113){\special{em:moveto}}
\put(1031,133){\special{em:lineto}}
\put(1031,832){\special{em:moveto}}
\put(1031,812){\special{em:lineto}}
\put(1166,113){\special{em:moveto}}
\put(1166,133){\special{em:lineto}}
\put(1166,832){\special{em:moveto}}
\put(1166,812){\special{em:lineto}}
\put(1166,68){\makebox(0,0){0.06}}
\put(1301,113){\special{em:moveto}}
\put(1301,133){\special{em:lineto}}
\put(1301,832){\special{em:moveto}}
\put(1301,812){\special{em:lineto}}
\put(1436,113){\special{em:moveto}}
\put(1436,133){\special{em:lineto}}
\put(1436,832){\special{em:moveto}}
\put(1436,812){\special{em:lineto}}
\put(220,113){\special{em:moveto}}
\put(1436,113){\special{em:lineto}}
\put(1436,832){\special{em:lineto}}
\put(220,832){\special{em:lineto}}
\put(220,113){\special{em:lineto}}
 \put(-30,472){\makebox(0,0){$\phi^\#$}}
 \put(828,-40){\makebox(0,0){$1/k_q-1/k_{c}$}}
\put(342,199){\raisebox{-.8pt}{\makebox(0,0){$\Diamond$}}}
\put(1315,626){\raisebox{-.8pt}{\makebox(0,0){$\Diamond$}}}
\put(1108,532){\raisebox{-.8pt}{\makebox(0,0){$\Diamond$}}}
\put(903,443){\raisebox{-.8pt}{\makebox(0,0){$\Diamond$}}}
\put(698,357){\raisebox{-.8pt}{\makebox(0,0){$\Diamond$}}}
\put(342,120){\special{em:moveto}}
\put(342,278){\special{em:lineto}}
\put(332,120){\special{em:moveto}}
\put(352,120){\special{em:lineto}}
\put(332,278){\special{em:moveto}}
\put(352,278){\special{em:lineto}}
\put(1315,565){\special{em:moveto}}
\put(1315,687){\special{em:lineto}}
\put(1305,565){\special{em:moveto}}
\put(1325,565){\special{em:lineto}}
\put(1305,687){\special{em:moveto}}
\put(1325,687){\special{em:lineto}}
\put(1108,474){\special{em:moveto}}
\put(1108,590){\special{em:lineto}}
\put(1098,474){\special{em:moveto}}
\put(1118,474){\special{em:lineto}}
\put(1098,590){\special{em:moveto}}
\put(1118,590){\special{em:lineto}}
\put(903,369){\special{em:moveto}}
\put(903,517){\special{em:lineto}}
\put(893,369){\special{em:moveto}}
\put(913,369){\special{em:lineto}}
\put(893,517){\special{em:moveto}}
\put(913,517){\special{em:lineto}}
\put(698,286){\special{em:moveto}}
\put(698,428){\special{em:lineto}}
\put(688,286){\special{em:moveto}}
\put(708,286){\special{em:lineto}}
\put(688,428){\special{em:moveto}}
\put(708,428){\special{em:lineto}}
\multiput(220,140)(19.019,8.312){8}{\usebox{\plotpoint}}
\multiput(355,199)(19.021,8.305){56}{\usebox{\plotpoint}}
\put(1436,671){\usebox{\plotpoint}}
\sbox{\plotpoint}{\rule[-0.400pt]{0.800pt}{0.800pt}}%
\special{em:linewidth 0.8pt}%
\put(355,333){\makebox(0,0){$+$}}
\put(1328,642){\makebox(0,0){$+$}}
\put(1122,576){\makebox(0,0){$+$}}
\put(916,510){\makebox(0,0){$+$}}
\put(712,447){\makebox(0,0){$+$}}
\put(355,306){\special{em:moveto}}
\put(355,359){\special{em:lineto}}
\put(345,306){\special{em:moveto}}
\put(365,306){\special{em:lineto}}
\put(345,359){\special{em:moveto}}
\put(365,359){\special{em:lineto}}
\put(1328,616){\special{em:moveto}}
\put(1328,668){\special{em:lineto}}
\put(1318,616){\special{em:moveto}}
\put(1338,616){\special{em:lineto}}
\put(1318,668){\special{em:moveto}}
\put(1338,668){\special{em:lineto}}
\put(1122,550){\special{em:moveto}}
\put(1122,602){\special{em:lineto}}
\put(1112,550){\special{em:moveto}}
\put(1132,550){\special{em:lineto}}
\put(1112,602){\special{em:moveto}}
\put(1132,602){\special{em:lineto}}
\put(916,484){\special{em:moveto}}
\put(916,535){\special{em:lineto}}
\put(906,484){\special{em:moveto}}
\put(926,484){\special{em:lineto}}
\put(906,535){\special{em:moveto}}
\put(926,535){\special{em:lineto}}
\put(712,421){\special{em:moveto}}
\put(712,473){\special{em:lineto}}
\put(702,421){\special{em:moveto}}
\put(722,421){\special{em:lineto}}
\put(702,473){\special{em:moveto}}
\put(722,473){\special{em:lineto}}
\sbox{\plotpoint}{\rule[-0.500pt]{1.000pt}{1.000pt}}%
\special{em:linewidth 1.0pt}%
\multiput(220,290)(19.777,6.299){7}{\usebox{\plotpoint}}
\multiput(355,333)(19.783,6.277){55}{\usebox{\plotpoint}}
\put(1436,676){\usebox{\plotpoint}}
\sbox{\plotpoint}{\rule[-0.600pt]{1.200pt}{1.200pt}}%
\special{em:linewidth 1.2pt}%
\put(355,218){\raisebox{-.8pt}{\makebox(0,0){$\Box$}}}
\put(1328,502){\raisebox{-.8pt}{\makebox(0,0){$\Box$}}}
\put(1122,441){\raisebox{-.8pt}{\makebox(0,0){$\Box$}}}
\put(916,380){\raisebox{-.8pt}{\makebox(0,0){$\Box$}}}
\put(712,323){\raisebox{-.8pt}{\makebox(0,0){$\Box$}}}
\put(355,192){\special{em:moveto}}
\put(355,244){\special{em:lineto}}
\put(345,192){\special{em:moveto}}
\put(365,192){\special{em:lineto}}
\put(345,244){\special{em:moveto}}
\put(365,244){\special{em:lineto}}
\put(1328,477){\special{em:moveto}}
\put(1328,527){\special{em:lineto}}
\put(1318,477){\special{em:moveto}}
\put(1338,477){\special{em:lineto}}
\put(1318,527){\special{em:moveto}}
\put(1338,527){\special{em:lineto}}
\put(1122,416){\special{em:moveto}}
\put(1122,466){\special{em:lineto}}
\put(1112,416){\special{em:moveto}}
\put(1132,416){\special{em:lineto}}
\put(1112,466){\special{em:moveto}}
\put(1132,466){\special{em:lineto}}
\put(916,355){\special{em:moveto}}
\put(916,405){\special{em:lineto}}
\put(906,355){\special{em:moveto}}
\put(926,355){\special{em:lineto}}
\put(906,405){\special{em:moveto}}
\put(926,405){\special{em:lineto}}
\put(712,297){\special{em:moveto}}
\put(712,348){\special{em:lineto}}
\put(702,297){\special{em:moveto}}
\put(722,297){\special{em:lineto}}
\put(702,348){\special{em:moveto}}
\put(722,348){\special{em:lineto}}
\sbox{\plotpoint}{\rule[-0.500pt]{1.000pt}{1.000pt}}%
\special{em:linewidth 1.0pt}%
\multiput(220,178)(39.801,11.793){4}{\usebox{\plotpoint}}
\multiput(355,218)(39.853,11.613){27}{\usebox{\plotpoint}}
\put(1436,533){\usebox{\plotpoint}}
\sbox{\plotpoint}{\rule[-0.200pt]{0.400pt}{0.400pt}}%
\special{em:linewidth 0.4pt}%
\put(355,113){\special{em:moveto}}
\put(355,832){\special{em:lineto}}
\end{picture}


\caption{APE results for $\phi^{\#}$ as a function of
the quark mass $1/\kappa_q - 1/\kappa_c$.
The data shown is for $\beta=6.4$ with the Clover action
for the light quarks.
The plot shows the results of the old method for $L_s=15$ (``+'')
and $17$ (``$\Box$''), and the new method (``$\Diamond$'')
(see text).
Points from the new method have been shifted horizontally
to aid clarity.}
\end{figure}


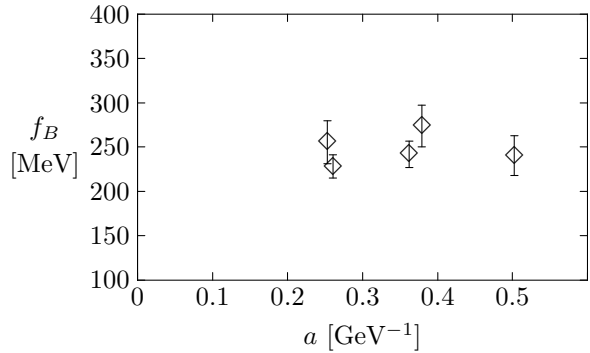
\begin{figure}
\protect\label{fig:fb_ape}

\setlength{\unitlength}{0.14pt}
\ifx\plotpoint\undefined\newsavebox{\plotpoint}\fi
\sbox{\plotpoint}{\rule[-0.200pt]{0.400pt}{0.400pt}}%
\special{em:linewidth 0.4pt}%
\begin{picture}(1500,900)(-80,0)
\font\gnuplot=cmr10 at 10pt
\gnuplot
\put(220,113){\special{em:moveto}}
\put(220,832){\special{em:lineto}}
\put(220,113){\special{em:moveto}}
\put(240,113){\special{em:lineto}}
\put(1436,113){\special{em:moveto}}
\put(1416,113){\special{em:lineto}}
\put(198,113){\makebox(0,0)[r]{100}}
\put(220,233){\special{em:moveto}}
\put(240,233){\special{em:lineto}}
\put(1436,233){\special{em:moveto}}
\put(1416,233){\special{em:lineto}}
\put(198,233){\makebox(0,0)[r]{150}}
\put(220,353){\special{em:moveto}}
\put(240,353){\special{em:lineto}}
\put(1436,353){\special{em:moveto}}
\put(1416,353){\special{em:lineto}}
\put(198,353){\makebox(0,0)[r]{200}}
\put(220,473){\special{em:moveto}}
\put(240,473){\special{em:lineto}}
\put(1436,473){\special{em:moveto}}
\put(1416,473){\special{em:lineto}}
\put(198,473){\makebox(0,0)[r]{250}}
\put(220,592){\special{em:moveto}}
\put(240,592){\special{em:lineto}}
\put(1436,592){\special{em:moveto}}
\put(1416,592){\special{em:lineto}}
\put(198,592){\makebox(0,0)[r]{300}}
\put(220,712){\special{em:moveto}}
\put(240,712){\special{em:lineto}}
\put(1436,712){\special{em:moveto}}
\put(1416,712){\special{em:lineto}}
\put(198,712){\makebox(0,0)[r]{350}}
\put(220,832){\special{em:moveto}}
\put(240,832){\special{em:lineto}}
\put(1436,832){\special{em:moveto}}
\put(1416,832){\special{em:lineto}}
\put(198,832){\makebox(0,0)[r]{400}}
\put(220,113){\special{em:moveto}}
\put(220,133){\special{em:lineto}}
\put(220,832){\special{em:moveto}}
\put(220,812){\special{em:lineto}}
\put(220,68){\makebox(0,0){0}}
\put(423,113){\special{em:moveto}}
\put(423,133){\special{em:lineto}}
\put(423,832){\special{em:moveto}}
\put(423,812){\special{em:lineto}}
\put(423,68){\makebox(0,0){0.1}}
\put(625,113){\special{em:moveto}}
\put(625,133){\special{em:lineto}}
\put(625,832){\special{em:moveto}}
\put(625,812){\special{em:lineto}}
\put(625,68){\makebox(0,0){0.2}}
\put(828,113){\special{em:moveto}}
\put(828,133){\special{em:lineto}}
\put(828,832){\special{em:moveto}}
\put(828,812){\special{em:lineto}}
\put(828,68){\makebox(0,0){0.3}}
\put(1031,113){\special{em:moveto}}
\put(1031,133){\special{em:lineto}}
\put(1031,832){\special{em:moveto}}
\put(1031,812){\special{em:lineto}}
\put(1031,68){\makebox(0,0){0.4}}
\put(1233,113){\special{em:moveto}}
\put(1233,133){\special{em:lineto}}
\put(1233,832){\special{em:moveto}}
\put(1233,812){\special{em:lineto}}
\put(1233,68){\makebox(0,0){0.5}}
\put(1436,113){\special{em:moveto}}
\put(1436,133){\special{em:lineto}}
\put(1436,832){\special{em:moveto}}
\put(1436,812){\special{em:lineto}}
\put(220,113){\special{em:moveto}}
\put(1436,113){\special{em:lineto}}
\put(1436,832){\special{em:lineto}}
\put(220,832){\special{em:lineto}}
\put(220,113){\special{em:lineto}}
 \put(-30,522){\makebox(0,0){$f_B$}}
 \put(-30,422){\makebox(0,0){[MeV]}}
 \put(828,-40){\makebox(0,0){$a$ [GeV$^{-1}$]}}
\put(1238,449){\raisebox{-.8pt}{\makebox(0,0){$\Diamond$}}}
\put(988,529){\raisebox{-.8pt}{\makebox(0,0){$\Diamond$}}}
\put(954,453){\raisebox{-.8pt}{\makebox(0,0){$\Diamond$}}}
\put(748,420){\raisebox{-.8pt}{\makebox(0,0){$\Diamond$}}}
\put(733,486){\raisebox{-.8pt}{\makebox(0,0){$\Diamond$}}}
\put(1238,396){\special{em:moveto}}
\put(1238,503){\special{em:lineto}}
\put(1228,396){\special{em:moveto}}
\put(1248,396){\special{em:lineto}}
\put(1228,503){\special{em:moveto}}
\put(1248,503){\special{em:lineto}}
\put(988,473){\special{em:moveto}}
\put(988,586){\special{em:lineto}}
\put(978,473){\special{em:moveto}}
\put(998,473){\special{em:lineto}}
\put(978,586){\special{em:moveto}}
\put(998,586){\special{em:lineto}}
\put(954,417){\special{em:moveto}}
\put(954,489){\special{em:lineto}}
\put(944,417){\special{em:moveto}}
\put(964,417){\special{em:lineto}}
\put(944,489){\special{em:moveto}}
\put(964,489){\special{em:lineto}}
\put(748,389){\special{em:moveto}}
\put(748,452){\special{em:lineto}}
\put(738,389){\special{em:moveto}}
\put(758,389){\special{em:lineto}}
\put(738,452){\special{em:moveto}}
\put(758,452){\special{em:lineto}}
\put(733,428){\special{em:moveto}}
\put(733,544){\special{em:lineto}}
\put(723,428){\special{em:moveto}}
\put(743,428){\special{em:lineto}}
\put(723,544){\special{em:moveto}}
\put(743,544){\special{em:lineto}}
\end{picture}


\caption{APE results for $\fbs$ for the various simulations
as a function of the lattice spacing $a$ (set from the rho mass).
From right to left the data is for Clover $\beta=6.0$,
Wilson $\beta=6.1$, Clover $\beta=6.2$, Wilson $\beta=6.4$
and Clover $\beta=6.4$.
Horizontal errors are suppressed for clarity.}
\end{figure}



\section{NRQCD Results}
\label{nrqcd}

The NRQCD action used in current simulations is correct to
$O(1/m_Q)$ \cite{nrqcd,beth}:
\[
{\cal L} = Q^\dagger \left\{ -D_t +
\frac{ \mbox{\boldmath $D$}^2}{2m_Q} +
\frac{g}{2m_Q} \mbox{\boldmath $\sigma . B$} \right\} Q.
\]
This action can be simply inverted to obtain the heavy quark propagator.
In addition to the $1/m_Q$ terms which appear in the action,
there are $1/m_Q$ terms which appear in the definition of
the currents \cite{lat94_arifa}. These must be included in order to make the
calculation correct at this order.
All the NRQCD work presented at this conference took these
terms into account.

The NRQCD approach is valid for heavy-light meson masses
down to around the B-meson. Unfortunately it is not valid
at the charm mass where higher order terms in the $1/m_Qa$
expansion become relevant \cite{lat94_john}. In theory, moving to a larger
$a$ value (smaller $\beta$) would mean that these terms
become less relevant. In practice however, this would only
increase the $O(m_q a)$ effects to an extent where the results
would be unreliable in any case.

Table \ref{tab:nrqcd} summarises the data
presented in this conference with the references appearing
in the second row.
The following sections detail the results of the SGO (SCRI-
Glasgow-Ohio) collaboration.
In the first, the quenched approximation was used, and in the
second dynamical quarks were included.
See \cite{lat95_terry} for details of the Kentucky calculation.

\begin{table*}[hbt]
\setlength{\tabcolsep}{1.5pc}
\caption{Summary of recent results using the NRQCD approach.
See text for detailed comments.
All numbers should be considered preliminary.}
\label{tab:nrqcd}
\begin{tabular*}{\textwidth}{@{}l@{\extracolsep{\fill}}ccc}
\hline
{\bf Collaboration} & SGO   & SGO    & Kentucky \\
Reference           &\cite{lat95_arifa}
                            & \cite{lat95_sara}
                                     & \cite{lat95_terry} \\
\hline
\multicolumn{4}{l}{{\bf $\!\!\!\!\!\!\!\!\!\!$Lattice Parameters}}  \\
$n_F$               & 0	    & 2      & 0      \\

$\beta$             & $6.0$ & $5.6$  & $6.0$  \\
Volume              & $16^3 \times 48$
                            & $16^3 \times 32$
                                     & $20^3 \times 30$ \\
$N_{cfgs}$          &$\sim 50$& $100$  & $32$   \\
$S_q$               & Clover& Wilson \& Clover& Wilson \\
$S_Q$               & NRQCD to $O(1/m_Q)$
                            & NRQCD to $O(1/m_Q)$
                                     & NRQCD to $O(1/m_Q)$ \\
                    & \& Static
                            & \& Static & \\
\hline
{\bf Results}       &       &        & \\
$\fb$ [MeV]         & $\sim160(40)$
                            & $\sim190$
                                     & \\
$f_{B_S}^{stat}/\fbs$
                    &       & $1.18(3)$ Clover
                                     & $1.15^{+2}_{-1}$ \\
                    &       & $1.28(4)$ Wilson
                                     & \\
\hline
\end{tabular*}
\end{table*}

\subsection{SGO Results: Quenched}

The SGO collaboration's quenched data \cite{lat95_arifa}
are run using the Clover action for the light quarks:
both the tadpole unimproved and improved (i.e. $c=1$ and $\sim 1.4$
respectively, where $c$ is the coefficient of the Clover term in the action).
Two values of $\kappa_q$ were used which straddle
the mass of the strange quark.
Four values of the NRQCD quark mass were used.
The results for $\phi^{\#}$ at the chiral limit versus $1/M_P^\#$
are plotted in fig.4 
for the $c=1$ case together with the static point.
Fitting only the NRQCD points shows that they can
be made to extrapolate to the static point only if a quadratic
term, $c_2/M_P^2$, is included in the fit.
If a linear term only is allowed, then the static
point seems too high.
However, it may be that with the inclusion of more
$m_Q$ values closer to the static point, this apparent discrepancy
disappears (see sec.\ref{sgo_dyn}).
Also, the $c=1.4$ data (not shown) has a better agreement between
the NRQCD data and the static point \cite{lat95_arifa}.

A preliminary calculation of the renormalisation constant
relevant for the NRQCD-light axial current has been performed
\cite{junko}. Using this value, the SGO collaboration
obtains $f_P$ at their 4 values of $m_Q$.
From these, I obtain the preliminary estimate
of $\fb$ which appears in table \ref{tab:nrqcd}.

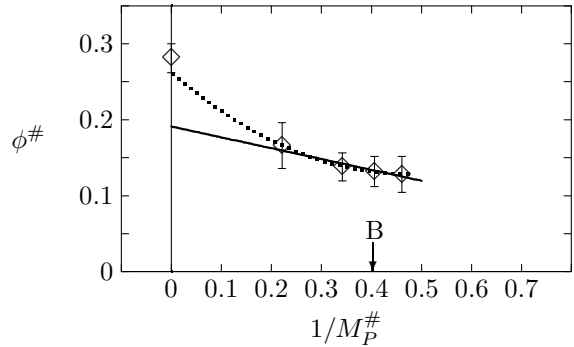
\begin{figure}[t]
\protect\label{fig:phi_arifa}.

\setlength{\unitlength}{0.14pt}
\ifx\plotpoint\undefined\newsavebox{\plotpoint}\fi
\sbox{\plotpoint}{\rule[-0.200pt]{0.400pt}{0.400pt}}%
\special{em:linewidth 0.4pt}%
\begin{picture}(1500,900)(-80,0)
\font\gnuplot=cmr10 at 10pt
\gnuplot

\put(900,193){\vector(0,-1){80}}
\put(880,200){B}

\put(220,113){\special{em:moveto}}
\put(1436,113){\special{em:lineto}}
\put(355,113){\special{em:moveto}}
\put(355,832){\special{em:lineto}}
\put(220,113){\special{em:moveto}}
\put(240,113){\special{em:lineto}}
\put(1436,113){\special{em:moveto}}
\put(1416,113){\special{em:lineto}}
\put(198,113){\makebox(0,0)[r]{0}}
\put(220,216){\special{em:moveto}}
\put(240,216){\special{em:lineto}}
\put(1436,216){\special{em:moveto}}
\put(1416,216){\special{em:lineto}}
\put(220,318){\special{em:moveto}}
\put(240,318){\special{em:lineto}}
\put(1436,318){\special{em:moveto}}
\put(1416,318){\special{em:lineto}}
\put(198,318){\makebox(0,0)[r]{0.1}}
\put(220,421){\special{em:moveto}}
\put(240,421){\special{em:lineto}}
\put(1436,421){\special{em:moveto}}
\put(1416,421){\special{em:lineto}}
\put(220,524){\special{em:moveto}}
\put(240,524){\special{em:lineto}}
\put(1436,524){\special{em:moveto}}
\put(1416,524){\special{em:lineto}}
\put(198,524){\makebox(0,0)[r]{0.2}}
\put(220,627){\special{em:moveto}}
\put(240,627){\special{em:lineto}}
\put(1436,627){\special{em:moveto}}
\put(1416,627){\special{em:lineto}}
\put(220,729){\special{em:moveto}}
\put(240,729){\special{em:lineto}}
\put(1436,729){\special{em:moveto}}
\put(1416,729){\special{em:lineto}}
\put(198,729){\makebox(0,0)[r]{0.3}}
\put(220,832){\special{em:moveto}}
\put(240,832){\special{em:lineto}}
\put(1436,832){\special{em:moveto}}
\put(1416,832){\special{em:lineto}}
\put(220,113){\special{em:moveto}}
\put(220,133){\special{em:lineto}}
\put(220,832){\special{em:moveto}}
\put(220,812){\special{em:lineto}}
\put(355,113){\special{em:moveto}}
\put(355,133){\special{em:lineto}}
\put(355,832){\special{em:moveto}}
\put(355,812){\special{em:lineto}}
\put(355,68){\makebox(0,0){0}}
\put(490,113){\special{em:moveto}}
\put(490,133){\special{em:lineto}}
\put(490,832){\special{em:moveto}}
\put(490,812){\special{em:lineto}}
\put(490,68){\makebox(0,0){0.1}}
\put(625,113){\special{em:moveto}}
\put(625,133){\special{em:lineto}}
\put(625,832){\special{em:moveto}}
\put(625,812){\special{em:lineto}}
\put(625,68){\makebox(0,0){0.2}}
\put(760,113){\special{em:moveto}}
\put(760,133){\special{em:lineto}}
\put(760,832){\special{em:moveto}}
\put(760,812){\special{em:lineto}}
\put(760,68){\makebox(0,0){0.3}}
\put(896,113){\special{em:moveto}}
\put(896,133){\special{em:lineto}}
\put(896,832){\special{em:moveto}}
\put(896,812){\special{em:lineto}}
\put(896,68){\makebox(0,0){0.4}}
\put(1031,113){\special{em:moveto}}
\put(1031,133){\special{em:lineto}}
\put(1031,832){\special{em:moveto}}
\put(1031,812){\special{em:lineto}}
\put(1031,68){\makebox(0,0){0.5}}
\put(1166,113){\special{em:moveto}}
\put(1166,133){\special{em:lineto}}
\put(1166,832){\special{em:moveto}}
\put(1166,812){\special{em:lineto}}
\put(1166,68){\makebox(0,0){0.6}}
\put(1301,113){\special{em:moveto}}
\put(1301,133){\special{em:lineto}}
\put(1301,832){\special{em:moveto}}
\put(1301,812){\special{em:lineto}}
\put(1301,68){\makebox(0,0){0.7}}
\put(1436,113){\special{em:moveto}}
\put(1436,133){\special{em:lineto}}
\put(1436,832){\special{em:moveto}}
\put(1436,812){\special{em:lineto}}
\put(220,113){\special{em:moveto}}
\put(1436,113){\special{em:lineto}}
\put(1436,832){\special{em:lineto}}
\put(220,832){\special{em:lineto}}
\put(220,113){\special{em:lineto}}
 \put(-30,472){\makebox(0,0){$\phi^\#$}}
 \put(828,-40){\makebox(0,0){$1/M_P^\#$}}
\put(978,376){\raisebox{-.8pt}{\makebox(0,0){$\Diamond$}}}
\put(904,384){\raisebox{-.8pt}{\makebox(0,0){$\Diamond$}}}
\put(817,397){\raisebox{-.8pt}{\makebox(0,0){$\Diamond$}}}
\put(654,454){\raisebox{-.8pt}{\makebox(0,0){$\Diamond$}}}
\put(355,690){\raisebox{-.8pt}{\makebox(0,0){$\Diamond$}}}
\put(978,328){\special{em:moveto}}
\put(978,425){\special{em:lineto}}
\put(968,328){\special{em:moveto}}
\put(988,328){\special{em:lineto}}
\put(968,425){\special{em:moveto}}
\put(988,425){\special{em:lineto}}
\put(904,343){\special{em:moveto}}
\put(904,425){\special{em:lineto}}
\put(894,343){\special{em:moveto}}
\put(914,343){\special{em:lineto}}
\put(894,425){\special{em:moveto}}
\put(914,425){\special{em:lineto}}
\put(817,359){\special{em:moveto}}
\put(817,434){\special{em:lineto}}
\put(807,359){\special{em:moveto}}
\put(827,359){\special{em:lineto}}
\put(807,434){\special{em:moveto}}
\put(827,434){\special{em:lineto}}
\put(654,392){\special{em:moveto}}
\put(654,516){\special{em:lineto}}
\put(644,392){\special{em:moveto}}
\put(664,392){\special{em:lineto}}
\put(644,516){\special{em:moveto}}
\put(664,516){\special{em:lineto}}
\put(355,651){\special{em:moveto}}
\put(355,729){\special{em:lineto}}
\put(345,651){\special{em:moveto}}
\put(365,651){\special{em:lineto}}
\put(345,729){\special{em:moveto}}
\put(365,729){\special{em:lineto}}
\put(355,113){\usebox{\plotpoint}}
\put(355,832){\usebox{\plotpoint}}
\sbox{\plotpoint}{\rule[-0.400pt]{0.800pt}{0.800pt}}%
\special{em:linewidth 0.8pt}%
\put(355,506){\special{em:moveto}}
\put(362,504){\special{em:lineto}}
\put(369,503){\special{em:lineto}}
\put(375,501){\special{em:lineto}}
\put(382,500){\special{em:lineto}}
\put(389,498){\special{em:lineto}}
\put(396,497){\special{em:lineto}}
\put(402,495){\special{em:lineto}}
\put(409,494){\special{em:lineto}}
\put(416,492){\special{em:lineto}}
\put(423,491){\special{em:lineto}}
\put(429,490){\special{em:lineto}}
\put(436,488){\special{em:lineto}}
\put(443,487){\special{em:lineto}}
\put(450,485){\special{em:lineto}}
\put(456,484){\special{em:lineto}}
\put(463,482){\special{em:lineto}}
\put(470,481){\special{em:lineto}}
\put(477,479){\special{em:lineto}}
\put(483,478){\special{em:lineto}}
\put(490,476){\special{em:lineto}}
\put(497,475){\special{em:lineto}}
\put(504,473){\special{em:lineto}}
\put(510,472){\special{em:lineto}}
\put(517,470){\special{em:lineto}}
\put(524,469){\special{em:lineto}}
\put(531,468){\special{em:lineto}}
\put(538,466){\special{em:lineto}}
\put(544,465){\special{em:lineto}}
\put(551,463){\special{em:lineto}}
\put(558,462){\special{em:lineto}}
\put(565,460){\special{em:lineto}}
\put(571,459){\special{em:lineto}}
\put(578,457){\special{em:lineto}}
\put(585,456){\special{em:lineto}}
\put(592,454){\special{em:lineto}}
\put(598,453){\special{em:lineto}}
\put(605,451){\special{em:lineto}}
\put(612,450){\special{em:lineto}}
\put(619,448){\special{em:lineto}}
\put(625,447){\special{em:lineto}}
\put(632,446){\special{em:lineto}}
\put(639,444){\special{em:lineto}}
\put(646,443){\special{em:lineto}}
\put(652,441){\special{em:lineto}}
\put(659,440){\special{em:lineto}}
\put(666,438){\special{em:lineto}}
\put(673,437){\special{em:lineto}}
\put(679,435){\special{em:lineto}}
\put(686,434){\special{em:lineto}}
\put(693,432){\special{em:lineto}}
\put(700,431){\special{em:lineto}}
\put(706,429){\special{em:lineto}}
\put(713,428){\special{em:lineto}}
\put(720,427){\special{em:lineto}}
\put(727,425){\special{em:lineto}}
\put(733,424){\special{em:lineto}}
\put(740,422){\special{em:lineto}}
\put(747,421){\special{em:lineto}}
\put(754,419){\special{em:lineto}}
\put(760,418){\special{em:lineto}}
\put(767,416){\special{em:lineto}}
\put(774,415){\special{em:lineto}}
\put(781,413){\special{em:lineto}}
\put(787,412){\special{em:lineto}}
\put(794,410){\special{em:lineto}}
\put(801,409){\special{em:lineto}}
\put(808,407){\special{em:lineto}}
\put(814,406){\special{em:lineto}}
\put(821,405){\special{em:lineto}}
\put(828,403){\special{em:lineto}}
\put(835,402){\special{em:lineto}}
\put(842,400){\special{em:lineto}}
\put(848,399){\special{em:lineto}}
\put(855,397){\special{em:lineto}}
\put(862,396){\special{em:lineto}}
\put(869,394){\special{em:lineto}}
\put(875,393){\special{em:lineto}}
\put(882,391){\special{em:lineto}}
\put(889,390){\special{em:lineto}}
\put(896,388){\special{em:lineto}}
\put(902,387){\special{em:lineto}}
\put(909,385){\special{em:lineto}}
\put(916,384){\special{em:lineto}}
\put(923,383){\special{em:lineto}}
\put(929,381){\special{em:lineto}}
\put(936,380){\special{em:lineto}}
\put(943,378){\special{em:lineto}}
\put(950,377){\special{em:lineto}}
\put(956,375){\special{em:lineto}}
\put(963,374){\special{em:lineto}}
\put(970,372){\special{em:lineto}}
\put(977,371){\special{em:lineto}}
\put(983,369){\special{em:lineto}}
\put(990,368){\special{em:lineto}}
\put(997,366){\special{em:lineto}}
\put(1004,365){\special{em:lineto}}
\put(1010,363){\special{em:lineto}}
\put(1017,362){\special{em:lineto}}
\put(1024,361){\special{em:lineto}}
\put(1031,359){\special{em:lineto}}
\sbox{\plotpoint}{\rule[-0.500pt]{1.000pt}{1.000pt}}%
\special{em:linewidth 1.0pt}%
\put(355,647){\usebox{\plotpoint}}
\put(355.00,647.00){\usebox{\plotpoint}}
\multiput(363,641)(15.620,-13.668){0}{\usebox{\plotpoint}}
\put(371.09,633.92){\usebox{\plotpoint}}
\put(387.53,621.31){\usebox{\plotpoint}}
\multiput(388,621)(15.620,-13.668){0}{\usebox{\plotpoint}}
\put(403.65,608.26){\usebox{\plotpoint}}
\multiput(404,608)(15.620,-13.668){0}{\usebox{\plotpoint}}
\put(419.75,595.19){\usebox{\plotpoint}}
\multiput(420,595)(16.604,-12.453){0}{\usebox{\plotpoint}}
\multiput(428,589)(16.604,-12.453){0}{\usebox{\plotpoint}}
\put(436.35,582.74){\usebox{\plotpoint}}
\multiput(444,577)(16.604,-12.453){0}{\usebox{\plotpoint}}
\put(452.96,570.28){\usebox{\plotpoint}}
\multiput(460,565)(17.270,-11.513){0}{\usebox{\plotpoint}}
\put(469.96,558.40){\usebox{\plotpoint}}
\multiput(477,554)(16.604,-12.453){0}{\usebox{\plotpoint}}
\put(487.08,546.70){\usebox{\plotpoint}}
\multiput(493,543)(16.604,-12.453){0}{\usebox{\plotpoint}}
\put(504.20,535.00){\usebox{\plotpoint}}
\multiput(509,532)(17.601,-11.000){0}{\usebox{\plotpoint}}
\put(521.80,524.00){\usebox{\plotpoint}}
\multiput(525,522)(17.601,-11.000){0}{\usebox{\plotpoint}}
\put(539.60,513.33){\usebox{\plotpoint}}
\multiput(542,512)(17.601,-11.000){0}{\usebox{\plotpoint}}
\put(557.27,502.45){\usebox{\plotpoint}}
\multiput(558,502)(18.564,-9.282){0}{\usebox{\plotpoint}}
\multiput(566,498)(17.601,-11.000){0}{\usebox{\plotpoint}}
\put(575.36,492.32){\usebox{\plotpoint}}
\multiput(582,489)(17.601,-11.000){0}{\usebox{\plotpoint}}
\put(593.49,482.26){\usebox{\plotpoint}}
\multiput(598,480)(18.564,-9.282){0}{\usebox{\plotpoint}}
\put(611.91,472.71){\usebox{\plotpoint}}
\multiput(615,471)(18.564,-9.282){0}{\usebox{\plotpoint}}
\put(630.41,463.30){\usebox{\plotpoint}}
\multiput(631,463)(19.434,-7.288){0}{\usebox{\plotpoint}}
\multiput(639,460)(18.564,-9.282){0}{\usebox{\plotpoint}}
\put(649.33,454.84){\usebox{\plotpoint}}
\multiput(655,452)(19.434,-7.288){0}{\usebox{\plotpoint}}
\put(668.25,446.37){\usebox{\plotpoint}}
\multiput(671,445)(19.434,-7.288){0}{\usebox{\plotpoint}}
\multiput(679,442)(18.564,-9.282){0}{\usebox{\plotpoint}}
\put(687.18,437.94){\usebox{\plotpoint}}
\multiput(696,435)(19.434,-7.288){0}{\usebox{\plotpoint}}
\put(706.73,430.97){\usebox{\plotpoint}}
\multiput(712,429)(19.434,-7.288){0}{\usebox{\plotpoint}}
\put(726.17,423.69){\usebox{\plotpoint}}
\multiput(728,423)(19.434,-7.288){0}{\usebox{\plotpoint}}
\multiput(736,420)(19.434,-7.288){0}{\usebox{\plotpoint}}
\put(745.66,416.59){\usebox{\plotpoint}}
\multiput(752,415)(19.434,-7.288){0}{\usebox{\plotpoint}}
\put(765.54,410.77){\usebox{\plotpoint}}
\multiput(769,410)(20.136,-5.034){0}{\usebox{\plotpoint}}
\multiput(777,408)(19.434,-7.288){0}{\usebox{\plotpoint}}
\put(785.41,404.90){\usebox{\plotpoint}}
\multiput(793,403)(20.136,-5.034){0}{\usebox{\plotpoint}}
\put(805.54,399.86){\usebox{\plotpoint}}
\multiput(809,399)(20.136,-5.034){0}{\usebox{\plotpoint}}
\multiput(817,397)(20.136,-5.034){0}{\usebox{\plotpoint}}
\put(825.70,394.91){\usebox{\plotpoint}}
\multiput(833,394)(20.261,-4.503){0}{\usebox{\plotpoint}}
\put(846.05,390.99){\usebox{\plotpoint}}
\multiput(850,390)(20.595,-2.574){0}{\usebox{\plotpoint}}
\multiput(858,389)(20.136,-5.034){0}{\usebox{\plotpoint}}
\put(866.37,386.95){\usebox{\plotpoint}}
\multiput(874,386)(20.595,-2.574){0}{\usebox{\plotpoint}}
\put(886.97,384.38){\usebox{\plotpoint}}
\multiput(890,384)(20.595,-2.574){0}{\usebox{\plotpoint}}
\multiput(898,383)(20.595,-2.574){0}{\usebox{\plotpoint}}
\put(907.56,381.80){\usebox{\plotpoint}}
\multiput(914,381)(20.629,-2.292){0}{\usebox{\plotpoint}}
\put(928.21,380.00){\usebox{\plotpoint}}
\multiput(931,380)(20.595,-2.574){0}{\usebox{\plotpoint}}
\multiput(939,379)(20.756,0.000){0}{\usebox{\plotpoint}}
\put(948.89,378.76){\usebox{\plotpoint}}
\multiput(955,378)(20.756,0.000){0}{\usebox{\plotpoint}}
\put(969.60,378.00){\usebox{\plotpoint}}
\multiput(971,378)(20.595,-2.574){0}{\usebox{\plotpoint}}
\multiput(979,377)(20.756,0.000){0}{\usebox{\plotpoint}}
\put(990.29,377.00){\usebox{\plotpoint}}
\multiput(996,377)(20.595,2.574){0}{\usebox{\plotpoint}}
\end{picture}


\caption{Quenched SGO results for $\phi^{\#}$ as a function of
the inverse heavy-light meson mass $1/M_P\#$ at the chiral
limit [31]. 
The results of a linear and quadratic fit in $1/M_P^\#=1/(M_P a)$
to the NRQCD data points is also shown.}
\end{figure}

\subsection{SGO Results: Dynamical}
\label{sgo_dyn}

Dynamical simulations involving NRQCD were presented by the SGO
Collaboration in last year's conference \cite{lat94_sara}. This year's
results \cite{lat95_sara} are enhanced by the inclusion of more values
of $m_Q$, by a further analysis of the $O(1/m_Q)$ terms that contribute
to $f_P$, and by the use of the tadpole improved Clover action for
the light quarks.
Again the preliminary values for the renormalisation constant were used
\cite{junko}.
For the Wilson data only two values of the light quark mass were
available meaning that chiral extrapolations were not robust.
The Clover data has three light quark values.

In fig.5 
a plot of $\zren \phi^{\#}
=f_P \sqrt{M_P} a^{3/2}$ (see eq.(\ref{eq:frm_lat}))
versus $1/M_P^\#$ is displayed.
In this graph both Wilson (at $\kappa_q = 0.1585$) and
Clover ($\kappa_q = 0.1385$) data are plotted.
The $\kappa_q$ values are chosen so that the corresponding
pion masses agree.
Note that in order to make the comparison, the appropriate
renormalisation factor for both actions has been included.
The curves are cubic fits in $1/M_P^\#$ to the NRQCD data only.
The static points are also plotted, but not included in the
fits.

There are three important points to be made:

\begin{itemize}
\item

The NRQCD data points smoothly extrapolate to the static point. Thus
the inclusion of more data points with larger $m_Q$ seems to have
resolved the discrepancy that appeared earlier \cite{lat94_sara}.

\item
The slope of $\phi$ against $1/M_P$ is apparently increasing
as $M_P \rightarrow \infty$.
This signifies the presence of terms of order $1/m_Q^2$.
However, the calculation is correct only to $O(1/m_Q)$.
Therefore, strictly speaking, any non-constant behaviour
of this slope at finite $m_Q$ is not a real prediction
of these calculations.
In practice though, the behaviour seen in fig 5 
matches well with data from conventional simulations
at around the D-meson (see sec. \ref{conv}).
So presumably it true is that the present simulation
does not suffer from neglecting terms of order $1/m_Q^2$.
Note also that from the figure there is a 30\%
$1/M_P$ correction to $\fb$ compared with the static
value.

\item
The Wilson data is statistically lower than the Clover data.
The obvious explanation is that this is a symptom of
$O(a)$ effects spoiling the Wilson data \cite{cra}.
(These systematics could enter either in $\phi^{\#}$
or in $\ia$.)
It could also be that the one-loop perturbative calculation
of $\zren$ is inadequate, and that a non-perturbatively
defined $\zren$ would remove the disagreement.
\end{itemize}

In the SGO analysis the contributions of the three $O(1/m_Q)$ terms
to $\phi$ were extracted. Expressing these as $c_i/M_P$ they found that

\[
|c_{D^2}| > |c_{\sigma . D}| >> |c_{\sigma . B}|
\]
That the hyperfine component, $c_{\sigma . B}$ is smallest fits our
expectations since this term breaks {\em both} the spin {\em and}
flavour symmetry present in the $m_Q = \infty$ limit \cite{neubert,lat95_sara}.

Performing a chiral extrapolation and using a nominal value
of $\ia = 2$ GeV, one obtains the values for
$\fb$ and $\fb/f_{B_S}$ which appear in table \ref{tab:nrqcd}, \cite{lat95_sara}.
This (rough) $\fb$ value is in nice agreement with that obtained by
the quenched NRQCD analysis of \cite{lat95_arifa} implying that
the effects of dynamical quarks are at, or below, the level of statistics.
The difference between the two actions for the chiral ratio
is unexpected since systematic effects such as $O(a)$ are
expected to cancel in this ratio.

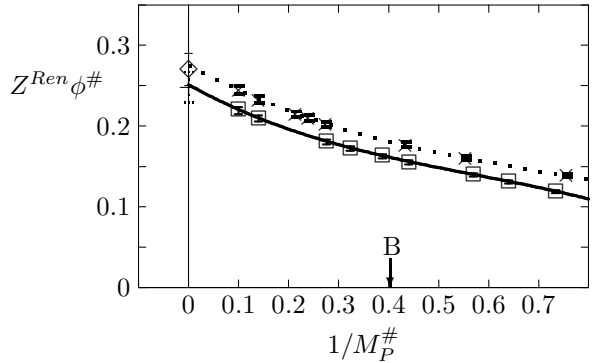
\begin{figure}[t]
\protect\label{fig:phi_sara}.

\setlength{\unitlength}{0.14pt}
\ifx\plotpoint\undefined\newsavebox{\plotpoint}\fi
\sbox{\plotpoint}{\rule[-0.200pt]{0.400pt}{0.400pt}}%
\special{em:linewidth 0.4pt}%
\begin{picture}(1500,900)(-80,0)
\font\gnuplot=cmr10 at 10pt
\gnuplot

\put(900,193){\vector(0,-1){80}}
\put(880,200){B}

\put(220,113){\special{em:moveto}}
\put(1436,113){\special{em:lineto}}
\put(355,113){\special{em:moveto}}
\put(355,877){\special{em:lineto}}
\put(220,113){\special{em:moveto}}
\put(240,113){\special{em:lineto}}
\put(1436,113){\special{em:moveto}}
\put(1416,113){\special{em:lineto}}
\put(198,113){\makebox(0,0)[r]{0}}
\put(220,222){\special{em:moveto}}
\put(240,222){\special{em:lineto}}
\put(1436,222){\special{em:moveto}}
\put(1416,222){\special{em:lineto}}
\put(220,331){\special{em:moveto}}
\put(240,331){\special{em:lineto}}
\put(1436,331){\special{em:moveto}}
\put(1416,331){\special{em:lineto}}
\put(198,331){\makebox(0,0)[r]{0.1}}
\put(220,440){\special{em:moveto}}
\put(240,440){\special{em:lineto}}
\put(1436,440){\special{em:moveto}}
\put(1416,440){\special{em:lineto}}
\put(220,550){\special{em:moveto}}
\put(240,550){\special{em:lineto}}
\put(1436,550){\special{em:moveto}}
\put(1416,550){\special{em:lineto}}
\put(198,550){\makebox(0,0)[r]{0.2}}
\put(220,659){\special{em:moveto}}
\put(240,659){\special{em:lineto}}
\put(1436,659){\special{em:moveto}}
\put(1416,659){\special{em:lineto}}
\put(220,768){\special{em:moveto}}
\put(240,768){\special{em:lineto}}
\put(1436,768){\special{em:moveto}}
\put(1416,768){\special{em:lineto}}
\put(198,768){\makebox(0,0)[r]{0.3}}
\put(220,877){\special{em:moveto}}
\put(240,877){\special{em:lineto}}
\put(1436,877){\special{em:moveto}}
\put(1416,877){\special{em:lineto}}
\put(220,113){\special{em:moveto}}
\put(220,133){\special{em:lineto}}
\put(220,877){\special{em:moveto}}
\put(220,857){\special{em:lineto}}
\put(355,113){\special{em:moveto}}
\put(355,133){\special{em:lineto}}
\put(355,877){\special{em:moveto}}
\put(355,857){\special{em:lineto}}
\put(355,68){\makebox(0,0){0}}
\put(490,113){\special{em:moveto}}
\put(490,133){\special{em:lineto}}
\put(490,877){\special{em:moveto}}
\put(490,857){\special{em:lineto}}
\put(490,68){\makebox(0,0){0.1}}
\put(625,113){\special{em:moveto}}
\put(625,133){\special{em:lineto}}
\put(625,877){\special{em:moveto}}
\put(625,857){\special{em:lineto}}
\put(625,68){\makebox(0,0){0.2}}
\put(760,113){\special{em:moveto}}
\put(760,133){\special{em:lineto}}
\put(760,877){\special{em:moveto}}
\put(760,857){\special{em:lineto}}
\put(760,68){\makebox(0,0){0.3}}
\put(896,113){\special{em:moveto}}
\put(896,133){\special{em:lineto}}
\put(896,877){\special{em:moveto}}
\put(896,857){\special{em:lineto}}
\put(896,68){\makebox(0,0){0.4}}
\put(1031,113){\special{em:moveto}}
\put(1031,133){\special{em:lineto}}
\put(1031,877){\special{em:moveto}}
\put(1031,857){\special{em:lineto}}
\put(1031,68){\makebox(0,0){0.5}}
\put(1166,113){\special{em:moveto}}
\put(1166,133){\special{em:lineto}}
\put(1166,877){\special{em:moveto}}
\put(1166,857){\special{em:lineto}}
\put(1166,68){\makebox(0,0){0.6}}
\put(1301,113){\special{em:moveto}}
\put(1301,133){\special{em:lineto}}
\put(1301,877){\special{em:moveto}}
\put(1301,857){\special{em:lineto}}
\put(1301,68){\makebox(0,0){0.7}}
\put(1436,113){\special{em:moveto}}
\put(1436,133){\special{em:lineto}}
\put(1436,877){\special{em:moveto}}
\put(1436,857){\special{em:lineto}}
\put(220,113){\special{em:moveto}}
\put(1436,113){\special{em:lineto}}
\put(1436,877){\special{em:lineto}}
\put(220,877){\special{em:lineto}}
\put(220,113){\special{em:lineto}}
 \put(0,660){\makebox(0,0){$Z^{Ren} \phi^\#$}}
 \put(828,-40){\makebox(0,0){$1/M_P^\#$}}
\put(355,702){\raisebox{-.8pt}{\makebox(0,0){$\Diamond$}}}
\put(355,659){\special{em:moveto}}
\put(355,746){\special{em:lineto}}
\put(345,659){\special{em:moveto}}
\put(365,659){\special{em:lineto}}
\put(345,746){\special{em:moveto}}
\put(365,746){\special{em:lineto}}
\put(355,654){\makebox(0,0){$+$}}
\multiput(355,613)(0.000,20.756){4}{\usebox{\plotpoint}}
\put(355,696){\usebox{\plotpoint}}
\put(345.00,613.00){\usebox{\plotpoint}}
\put(365,613){\usebox{\plotpoint}}
\put(345.00,696.00){\usebox{\plotpoint}}
\put(365,696){\usebox{\plotpoint}}
\sbox{\plotpoint}{\rule[-0.400pt]{0.800pt}{0.800pt}}%
\special{em:linewidth 0.8pt}%
\put(490,592){\raisebox{-.8pt}{\makebox(0,0){$\Box$}}}
\put(544,570){\raisebox{-.8pt}{\makebox(0,0){$\Box$}}}
\put(728,506){\raisebox{-.8pt}{\makebox(0,0){$\Box$}}}
\put(792,489){\raisebox{-.8pt}{\makebox(0,0){$\Box$}}}
\put(879,468){\raisebox{-.8pt}{\makebox(0,0){$\Box$}}}
\put(950,451){\raisebox{-.8pt}{\makebox(0,0){$\Box$}}}
\put(1125,417){\raisebox{-.8pt}{\makebox(0,0){$\Box$}}}
\put(1220,398){\raisebox{-.8pt}{\makebox(0,0){$\Box$}}}
\put(1347,372){\raisebox{-.8pt}{\makebox(0,0){$\Box$}}}
\put(490,582){\special{em:moveto}}
\put(490,601){\special{em:lineto}}
\put(480,582){\special{em:moveto}}
\put(500,582){\special{em:lineto}}
\put(480,601){\special{em:moveto}}
\put(500,601){\special{em:lineto}}
\put(544,562){\special{em:moveto}}
\put(544,578){\special{em:lineto}}
\put(534,562){\special{em:moveto}}
\put(554,562){\special{em:lineto}}
\put(534,578){\special{em:moveto}}
\put(554,578){\special{em:lineto}}
\put(728,500){\special{em:moveto}}
\put(728,513){\special{em:lineto}}
\put(718,500){\special{em:moveto}}
\put(738,500){\special{em:lineto}}
\put(718,513){\special{em:moveto}}
\put(738,513){\special{em:lineto}}
\put(792,483){\special{em:moveto}}
\put(792,495){\special{em:lineto}}
\put(782,483){\special{em:moveto}}
\put(802,483){\special{em:lineto}}
\put(782,495){\special{em:moveto}}
\put(802,495){\special{em:lineto}}
\put(879,462){\special{em:moveto}}
\put(879,473){\special{em:lineto}}
\put(869,462){\special{em:moveto}}
\put(889,462){\special{em:lineto}}
\put(869,473){\special{em:moveto}}
\put(889,473){\special{em:lineto}}
\put(950,446){\special{em:moveto}}
\put(950,457){\special{em:lineto}}
\put(940,446){\special{em:moveto}}
\put(960,446){\special{em:lineto}}
\put(940,457){\special{em:moveto}}
\put(960,457){\special{em:lineto}}
\put(1125,413){\special{em:moveto}}
\put(1125,422){\special{em:lineto}}
\put(1115,413){\special{em:moveto}}
\put(1135,413){\special{em:lineto}}
\put(1115,422){\special{em:moveto}}
\put(1135,422){\special{em:lineto}}
\put(1220,394){\special{em:moveto}}
\put(1220,403){\special{em:lineto}}
\put(1210,394){\special{em:moveto}}
\put(1230,394){\special{em:lineto}}
\put(1210,403){\special{em:moveto}}
\put(1230,403){\special{em:lineto}}
\put(1347,368){\special{em:moveto}}
\put(1347,376){\special{em:lineto}}
\put(1337,368){\special{em:moveto}}
\put(1357,368){\special{em:lineto}}
\put(1337,376){\special{em:moveto}}
\put(1357,376){\special{em:lineto}}
\sbox{\plotpoint}{\rule[-0.500pt]{1.000pt}{1.000pt}}%
\special{em:linewidth 1.0pt}%
\put(490,647){\makebox(0,0){$\times$}}
\put(543,620){\makebox(0,0){$\times$}}
\put(643,582){\makebox(0,0){$\times$}}
\put(677,570){\makebox(0,0){$\times$}}
\put(724,554){\makebox(0,0){$\times$}}
\put(940,498){\makebox(0,0){$\times$}}
\put(1102,463){\makebox(0,0){$\times$}}
\put(1375,416){\makebox(0,0){$\times$}}
\put(490.00,637.00){\usebox{\plotpoint}}
\put(490,657){\usebox{\plotpoint}}
\put(480.00,637.00){\usebox{\plotpoint}}
\put(500,637){\usebox{\plotpoint}}
\put(480.00,657.00){\usebox{\plotpoint}}
\put(500,657){\usebox{\plotpoint}}
\put(543.00,611.00){\usebox{\plotpoint}}
\put(543,630){\usebox{\plotpoint}}
\put(533.00,611.00){\usebox{\plotpoint}}
\put(553,611){\usebox{\plotpoint}}
\put(533.00,630.00){\usebox{\plotpoint}}
\put(553,630){\usebox{\plotpoint}}
\put(643.00,574.00){\usebox{\plotpoint}}
\put(643,589){\usebox{\plotpoint}}
\put(633.00,574.00){\usebox{\plotpoint}}
\put(653,574){\usebox{\plotpoint}}
\put(633.00,589.00){\usebox{\plotpoint}}
\put(653,589){\usebox{\plotpoint}}
\put(677.00,563.00){\usebox{\plotpoint}}
\put(677,578){\usebox{\plotpoint}}
\put(667.00,563.00){\usebox{\plotpoint}}
\put(687,563){\usebox{\plotpoint}}
\put(667.00,578.00){\usebox{\plotpoint}}
\put(687,578){\usebox{\plotpoint}}
\put(724.00,547.00){\usebox{\plotpoint}}
\put(724,561){\usebox{\plotpoint}}
\put(714.00,547.00){\usebox{\plotpoint}}
\put(734,547){\usebox{\plotpoint}}
\put(714.00,561.00){\usebox{\plotpoint}}
\put(734,561){\usebox{\plotpoint}}
\put(940.00,492.00){\usebox{\plotpoint}}
\put(940,505){\usebox{\plotpoint}}
\put(930.00,492.00){\usebox{\plotpoint}}
\put(950,492){\usebox{\plotpoint}}
\put(930.00,505.00){\usebox{\plotpoint}}
\put(950,505){\usebox{\plotpoint}}
\put(1102.00,457.00){\usebox{\plotpoint}}
\put(1102,469){\usebox{\plotpoint}}
\put(1092.00,457.00){\usebox{\plotpoint}}
\put(1112,457){\usebox{\plotpoint}}
\put(1092.00,469.00){\usebox{\plotpoint}}
\put(1112,469){\usebox{\plotpoint}}
\put(1375.00,411.00){\usebox{\plotpoint}}
\put(1375,421){\usebox{\plotpoint}}
\put(1365.00,411.00){\usebox{\plotpoint}}
\put(1385,411){\usebox{\plotpoint}}
\put(1365.00,421.00){\usebox{\plotpoint}}
\put(1385,421){\usebox{\plotpoint}}
\sbox{\plotpoint}{\rule[-0.600pt]{1.200pt}{1.200pt}}%
\special{em:linewidth 1.2pt}%
\put(355,661){\special{em:moveto}}
\put(361,658){\special{em:lineto}}
\put(366,655){\special{em:lineto}}
\put(371,652){\special{em:lineto}}
\put(377,650){\special{em:lineto}}
\put(382,647){\special{em:lineto}}
\put(388,644){\special{em:lineto}}
\put(393,641){\special{em:lineto}}
\put(399,638){\special{em:lineto}}
\put(404,635){\special{em:lineto}}
\put(410,632){\special{em:lineto}}
\put(415,630){\special{em:lineto}}
\put(421,627){\special{em:lineto}}
\put(426,624){\special{em:lineto}}
\put(431,622){\special{em:lineto}}
\put(437,619){\special{em:lineto}}
\put(442,616){\special{em:lineto}}
\put(448,614){\special{em:lineto}}
\put(453,611){\special{em:lineto}}
\put(459,609){\special{em:lineto}}
\put(464,606){\special{em:lineto}}
\put(470,603){\special{em:lineto}}
\put(475,601){\special{em:lineto}}
\put(481,599){\special{em:lineto}}
\put(486,596){\special{em:lineto}}
\put(491,594){\special{em:lineto}}
\put(497,591){\special{em:lineto}}
\put(502,589){\special{em:lineto}}
\put(508,587){\special{em:lineto}}
\put(513,584){\special{em:lineto}}
\put(519,582){\special{em:lineto}}
\put(524,580){\special{em:lineto}}
\put(530,577){\special{em:lineto}}
\put(535,575){\special{em:lineto}}
\put(541,573){\special{em:lineto}}
\put(546,571){\special{em:lineto}}
\put(551,569){\special{em:lineto}}
\put(557,566){\special{em:lineto}}
\put(562,564){\special{em:lineto}}
\put(568,562){\special{em:lineto}}
\put(573,560){\special{em:lineto}}
\put(579,558){\special{em:lineto}}
\put(584,556){\special{em:lineto}}
\put(590,554){\special{em:lineto}}
\put(595,552){\special{em:lineto}}
\put(601,550){\special{em:lineto}}
\put(606,548){\special{em:lineto}}
\put(611,546){\special{em:lineto}}
\put(617,544){\special{em:lineto}}
\put(622,542){\special{em:lineto}}
\put(628,540){\special{em:lineto}}
\put(633,538){\special{em:lineto}}
\put(639,537){\special{em:lineto}}
\put(644,535){\special{em:lineto}}
\put(650,533){\special{em:lineto}}
\put(655,531){\special{em:lineto}}
\put(661,529){\special{em:lineto}}
\put(666,528){\special{em:lineto}}
\put(671,526){\special{em:lineto}}
\put(677,524){\special{em:lineto}}
\put(682,522){\special{em:lineto}}
\put(688,521){\special{em:lineto}}
\put(693,519){\special{em:lineto}}
\put(699,517){\special{em:lineto}}
\put(704,516){\special{em:lineto}}
\put(710,514){\special{em:lineto}}
\put(715,512){\special{em:lineto}}
\put(721,511){\special{em:lineto}}
\put(726,509){\special{em:lineto}}
\put(731,507){\special{em:lineto}}
\put(737,506){\special{em:lineto}}
\put(742,504){\special{em:lineto}}
\put(748,503){\special{em:lineto}}
\put(753,501){\special{em:lineto}}
\put(759,500){\special{em:lineto}}
\put(764,498){\special{em:lineto}}
\put(770,497){\special{em:lineto}}
\put(775,495){\special{em:lineto}}
\put(781,494){\special{em:lineto}}
\put(786,492){\special{em:lineto}}
\put(791,491){\special{em:lineto}}
\put(797,490){\special{em:lineto}}
\put(802,488){\special{em:lineto}}
\put(808,487){\special{em:lineto}}
\put(813,485){\special{em:lineto}}
\put(819,484){\special{em:lineto}}
\put(824,483){\special{em:lineto}}
\put(830,481){\special{em:lineto}}
\put(835,480){\special{em:lineto}}
\put(841,479){\special{em:lineto}}
\put(846,477){\special{em:lineto}}
\put(851,476){\special{em:lineto}}
\put(857,475){\special{em:lineto}}
\put(862,473){\special{em:lineto}}
\put(868,472){\special{em:lineto}}
\put(873,471){\special{em:lineto}}
\put(879,469){\special{em:lineto}}
\put(884,468){\special{em:lineto}}
\put(890,467){\special{em:lineto}}
\put(895,466){\special{em:lineto}}
\put(901,465){\special{em:lineto}}
\put(906,463){\special{em:lineto}}
\put(911,462){\special{em:lineto}}
\put(917,461){\special{em:lineto}}
\put(922,460){\special{em:lineto}}
\put(928,458){\special{em:lineto}}
\put(933,457){\special{em:lineto}}
\put(939,456){\special{em:lineto}}
\put(944,455){\special{em:lineto}}
\put(950,454){\special{em:lineto}}
\put(955,453){\special{em:lineto}}
\put(961,451){\special{em:lineto}}
\put(966,450){\special{em:lineto}}
\put(971,449){\special{em:lineto}}
\put(977,448){\special{em:lineto}}
\put(982,447){\special{em:lineto}}
\put(988,446){\special{em:lineto}}
\put(993,445){\special{em:lineto}}
\put(999,444){\special{em:lineto}}
\put(1004,442){\special{em:lineto}}
\put(1010,441){\special{em:lineto}}
\put(1015,440){\special{em:lineto}}
\put(1021,439){\special{em:lineto}}
\put(1026,438){\special{em:lineto}}
\put(1031,437){\special{em:lineto}}
\put(1037,436){\special{em:lineto}}
\put(1042,435){\special{em:lineto}}
\put(1048,434){\special{em:lineto}}
\put(1053,433){\special{em:lineto}}
\put(1059,432){\special{em:lineto}}
\put(1064,430){\special{em:lineto}}
\put(1070,429){\special{em:lineto}}
\put(1075,428){\special{em:lineto}}
\put(1081,427){\special{em:lineto}}
\put(1086,426){\special{em:lineto}}
\put(1091,425){\special{em:lineto}}
\put(1097,424){\special{em:lineto}}
\put(1102,423){\special{em:lineto}}
\put(1108,422){\special{em:lineto}}
\put(1113,421){\special{em:lineto}}
\put(1119,420){\special{em:lineto}}
\put(1124,419){\special{em:lineto}}
\put(1130,418){\special{em:lineto}}
\put(1135,417){\special{em:lineto}}
\put(1141,416){\special{em:lineto}}
\put(1146,415){\special{em:lineto}}
\put(1151,414){\special{em:lineto}}
\put(1157,412){\special{em:lineto}}
\put(1162,411){\special{em:lineto}}
\put(1168,410){\special{em:lineto}}
\put(1173,409){\special{em:lineto}}
\put(1179,408){\special{em:lineto}}
\put(1184,407){\special{em:lineto}}
\put(1190,406){\special{em:lineto}}
\put(1195,405){\special{em:lineto}}
\put(1201,404){\special{em:lineto}}
\put(1206,403){\special{em:lineto}}
\put(1211,402){\special{em:lineto}}
\put(1217,401){\special{em:lineto}}
\put(1222,400){\special{em:lineto}}
\put(1228,398){\special{em:lineto}}
\put(1233,397){\special{em:lineto}}
\put(1239,396){\special{em:lineto}}
\put(1244,395){\special{em:lineto}}
\put(1250,394){\special{em:lineto}}
\put(1255,393){\special{em:lineto}}
\put(1261,392){\special{em:lineto}}
\put(1266,391){\special{em:lineto}}
\put(1271,390){\special{em:lineto}}
\put(1277,389){\special{em:lineto}}
\put(1282,387){\special{em:lineto}}
\put(1288,386){\special{em:lineto}}
\put(1293,385){\special{em:lineto}}
\put(1299,384){\special{em:lineto}}
\put(1304,383){\special{em:lineto}}
\put(1310,382){\special{em:lineto}}
\put(1315,380){\special{em:lineto}}
\put(1321,379){\special{em:lineto}}
\put(1326,378){\special{em:lineto}}
\put(1331,377){\special{em:lineto}}
\put(1337,376){\special{em:lineto}}
\put(1342,374){\special{em:lineto}}
\put(1348,373){\special{em:lineto}}
\put(1353,372){\special{em:lineto}}
\put(1359,371){\special{em:lineto}}
\put(1364,370){\special{em:lineto}}
\put(1370,368){\special{em:lineto}}
\put(1375,367){\special{em:lineto}}
\put(1381,366){\special{em:lineto}}
\put(1386,364){\special{em:lineto}}
\put(1391,363){\special{em:lineto}}
\put(1397,362){\special{em:lineto}}
\put(1402,361){\special{em:lineto}}
\put(1408,359){\special{em:lineto}}
\put(1413,358){\special{em:lineto}}
\put(1419,357){\special{em:lineto}}
\put(1424,355){\special{em:lineto}}
\put(1430,354){\special{em:lineto}}
\put(1435,353){\special{em:lineto}}
\put(1436,352){\special{em:lineto}}
\sbox{\plotpoint}{\rule[-0.500pt]{1.000pt}{1.000pt}}%
\special{em:linewidth 1.0pt}%
\put(355,713){\usebox{\plotpoint}}
\put(355.00,713.00){\usebox{\plotpoint}}
\multiput(361,710)(35.595,-21.357){0}{\usebox{\plotpoint}}
\multiput(366,707)(37.129,-18.564){0}{\usebox{\plotpoint}}
\multiput(372,704)(37.129,-18.564){0}{\usebox{\plotpoint}}
\multiput(378,701)(35.595,-21.357){0}{\usebox{\plotpoint}}
\multiput(383,698)(37.129,-18.564){0}{\usebox{\plotpoint}}
\put(391.59,693.45){\usebox{\plotpoint}}
\multiput(394,692)(37.129,-18.564){0}{\usebox{\plotpoint}}
\multiput(400,689)(37.129,-18.564){0}{\usebox{\plotpoint}}
\multiput(406,686)(35.595,-21.357){0}{\usebox{\plotpoint}}
\multiput(411,683)(39.381,-13.127){0}{\usebox{\plotpoint}}
\multiput(417,681)(35.595,-21.357){0}{\usebox{\plotpoint}}
\multiput(422,678)(37.129,-18.564){0}{\usebox{\plotpoint}}
\put(428.52,674.74){\usebox{\plotpoint}}
\multiput(434,672)(35.595,-21.357){0}{\usebox{\plotpoint}}
\multiput(439,669)(39.381,-13.127){0}{\usebox{\plotpoint}}
\multiput(445,667)(35.595,-21.357){0}{\usebox{\plotpoint}}
\multiput(450,664)(37.129,-18.564){0}{\usebox{\plotpoint}}
\multiput(456,661)(39.381,-13.127){0}{\usebox{\plotpoint}}
\put(465.75,656.75){\usebox{\plotpoint}}
\multiput(467,656)(37.129,-18.564){0}{\usebox{\plotpoint}}
\multiput(473,653)(39.381,-13.127){0}{\usebox{\plotpoint}}
\multiput(479,651)(35.595,-21.357){0}{\usebox{\plotpoint}}
\multiput(484,648)(37.129,-18.564){0}{\usebox{\plotpoint}}
\multiput(490,645)(38.542,-15.417){0}{\usebox{\plotpoint}}
\multiput(495,643)(37.129,-18.564){0}{\usebox{\plotpoint}}
\put(503.26,639.25){\usebox{\plotpoint}}
\multiput(507,638)(35.595,-21.357){0}{\usebox{\plotpoint}}
\multiput(512,635)(39.381,-13.127){0}{\usebox{\plotpoint}}
\multiput(518,633)(35.595,-21.357){0}{\usebox{\plotpoint}}
\multiput(523,630)(39.381,-13.127){0}{\usebox{\plotpoint}}
\multiput(529,628)(39.381,-13.127){0}{\usebox{\plotpoint}}
\multiput(535,626)(35.595,-21.357){0}{\usebox{\plotpoint}}
\put(541.05,622.65){\usebox{\plotpoint}}
\multiput(546,621)(35.595,-21.357){0}{\usebox{\plotpoint}}
\multiput(551,618)(39.381,-13.127){0}{\usebox{\plotpoint}}
\multiput(557,616)(39.381,-13.127){0}{\usebox{\plotpoint}}
\multiput(563,614)(38.542,-15.417){0}{\usebox{\plotpoint}}
\multiput(568,612)(37.129,-18.564){0}{\usebox{\plotpoint}}
\put(579.42,607.19){\usebox{\plotpoint}}
\multiput(580,607)(38.542,-15.417){0}{\usebox{\plotpoint}}
\multiput(585,605)(39.381,-13.127){0}{\usebox{\plotpoint}}
\multiput(591,603)(35.595,-21.357){0}{\usebox{\plotpoint}}
\multiput(596,600)(39.381,-13.127){0}{\usebox{\plotpoint}}
\multiput(602,598)(39.381,-13.127){0}{\usebox{\plotpoint}}
\multiput(608,596)(38.542,-15.417){0}{\usebox{\plotpoint}}
\put(618.05,592.32){\usebox{\plotpoint}}
\multiput(619,592)(38.542,-15.417){0}{\usebox{\plotpoint}}
\multiput(624,590)(39.381,-13.127){0}{\usebox{\plotpoint}}
\multiput(630,588)(37.129,-18.564){0}{\usebox{\plotpoint}}
\multiput(636,585)(38.542,-15.417){0}{\usebox{\plotpoint}}
\multiput(641,583)(39.381,-13.127){0}{\usebox{\plotpoint}}
\multiput(647,581)(38.542,-15.417){0}{\usebox{\plotpoint}}
\put(656.74,577.42){\usebox{\plotpoint}}
\multiput(658,577)(39.381,-13.127){0}{\usebox{\plotpoint}}
\multiput(664,575)(38.542,-15.417){0}{\usebox{\plotpoint}}
\multiput(669,573)(39.381,-13.127){0}{\usebox{\plotpoint}}
\multiput(675,571)(40.946,-6.824){0}{\usebox{\plotpoint}}
\multiput(681,570)(38.542,-15.417){0}{\usebox{\plotpoint}}
\multiput(686,568)(39.381,-13.127){0}{\usebox{\plotpoint}}
\put(696.05,564.38){\usebox{\plotpoint}}
\multiput(697,564)(39.381,-13.127){0}{\usebox{\plotpoint}}
\multiput(703,562)(39.381,-13.127){0}{\usebox{\plotpoint}}
\multiput(709,560)(38.542,-15.417){0}{\usebox{\plotpoint}}
\multiput(714,558)(39.381,-13.127){0}{\usebox{\plotpoint}}
\multiput(720,556)(40.705,-8.141){0}{\usebox{\plotpoint}}
\multiput(725,555)(39.381,-13.127){0}{\usebox{\plotpoint}}
\put(735.46,551.51){\usebox{\plotpoint}}
\multiput(737,551)(38.542,-15.417){0}{\usebox{\plotpoint}}
\multiput(742,549)(40.946,-6.824){0}{\usebox{\plotpoint}}
\multiput(748,548)(38.542,-15.417){0}{\usebox{\plotpoint}}
\multiput(753,546)(39.381,-13.127){0}{\usebox{\plotpoint}}
\multiput(759,544)(39.381,-13.127){0}{\usebox{\plotpoint}}
\multiput(765,542)(40.705,-8.141){0}{\usebox{\plotpoint}}
\put(775.02,539.33){\usebox{\plotpoint}}
\multiput(776,539)(39.381,-13.127){0}{\usebox{\plotpoint}}
\multiput(782,537)(40.705,-8.141){0}{\usebox{\plotpoint}}
\multiput(787,536)(39.381,-13.127){0}{\usebox{\plotpoint}}
\multiput(793,534)(40.705,-8.141){0}{\usebox{\plotpoint}}
\multiput(798,533)(39.381,-13.127){0}{\usebox{\plotpoint}}
\multiput(804,531)(39.381,-13.127){0}{\usebox{\plotpoint}}
\put(814.88,528.02){\usebox{\plotpoint}}
\multiput(815,528)(39.381,-13.127){0}{\usebox{\plotpoint}}
\multiput(821,526)(40.705,-8.141){0}{\usebox{\plotpoint}}
\multiput(826,525)(39.381,-13.127){0}{\usebox{\plotpoint}}
\multiput(832,523)(40.946,-6.824){0}{\usebox{\plotpoint}}
\multiput(838,522)(38.542,-15.417){0}{\usebox{\plotpoint}}
\multiput(843,520)(40.946,-6.824){0}{\usebox{\plotpoint}}
\multiput(849,519)(38.542,-15.417){0}{\usebox{\plotpoint}}
\put(854.70,516.88){\usebox{\plotpoint}}
\multiput(860,516)(39.381,-13.127){0}{\usebox{\plotpoint}}
\multiput(866,514)(40.705,-8.141){0}{\usebox{\plotpoint}}
\multiput(871,513)(39.381,-13.127){0}{\usebox{\plotpoint}}
\multiput(877,511)(40.705,-8.141){0}{\usebox{\plotpoint}}
\multiput(882,510)(40.946,-6.824){0}{\usebox{\plotpoint}}
\multiput(888,509)(39.381,-13.127){0}{\usebox{\plotpoint}}
\put(894.86,506.83){\usebox{\plotpoint}}
\multiput(899,506)(39.381,-13.127){0}{\usebox{\plotpoint}}
\multiput(905,504)(40.946,-6.824){0}{\usebox{\plotpoint}}
\multiput(911,503)(40.705,-8.141){0}{\usebox{\plotpoint}}
\multiput(916,502)(39.381,-13.127){0}{\usebox{\plotpoint}}
\multiput(922,500)(40.705,-8.141){0}{\usebox{\plotpoint}}
\multiput(927,499)(40.946,-6.824){0}{\usebox{\plotpoint}}
\put(935.16,497.28){\usebox{\plotpoint}}
\multiput(939,496)(40.705,-8.141){0}{\usebox{\plotpoint}}
\multiput(944,495)(40.946,-6.824){0}{\usebox{\plotpoint}}
\multiput(950,494)(38.542,-15.417){0}{\usebox{\plotpoint}}
\multiput(955,492)(40.946,-6.824){0}{\usebox{\plotpoint}}
\multiput(961,491)(40.946,-6.824){0}{\usebox{\plotpoint}}
\multiput(967,490)(40.705,-8.141){0}{\usebox{\plotpoint}}
\put(975.45,487.85){\usebox{\plotpoint}}
\multiput(978,487)(40.705,-8.141){0}{\usebox{\plotpoint}}
\multiput(983,486)(40.946,-6.824){0}{\usebox{\plotpoint}}
\multiput(989,485)(40.946,-6.824){0}{\usebox{\plotpoint}}
\multiput(995,484)(40.705,-8.141){0}{\usebox{\plotpoint}}
\multiput(1000,483)(39.381,-13.127){0}{\usebox{\plotpoint}}
\multiput(1006,481)(40.946,-6.824){0}{\usebox{\plotpoint}}
\put(1015.97,479.21){\usebox{\plotpoint}}
\multiput(1017,479)(40.946,-6.824){0}{\usebox{\plotpoint}}
\multiput(1023,478)(40.705,-8.141){0}{\usebox{\plotpoint}}
\multiput(1028,477)(39.381,-13.127){0}{\usebox{\plotpoint}}
\multiput(1034,475)(40.946,-6.824){0}{\usebox{\plotpoint}}
\multiput(1040,474)(40.705,-8.141){0}{\usebox{\plotpoint}}
\multiput(1045,473)(40.946,-6.824){0}{\usebox{\plotpoint}}
\multiput(1051,472)(40.705,-8.141){0}{\usebox{\plotpoint}}
\put(1056.59,470.90){\usebox{\plotpoint}}
\multiput(1062,470)(40.946,-6.824){0}{\usebox{\plotpoint}}
\multiput(1068,469)(40.705,-8.141){0}{\usebox{\plotpoint}}
\multiput(1073,468)(39.381,-13.127){0}{\usebox{\plotpoint}}
\multiput(1079,466)(40.705,-8.141){0}{\usebox{\plotpoint}}
\multiput(1084,465)(40.946,-6.824){0}{\usebox{\plotpoint}}
\multiput(1090,464)(40.946,-6.824){0}{\usebox{\plotpoint}}
\put(1097.23,462.75){\usebox{\plotpoint}}
\multiput(1101,462)(40.946,-6.824){0}{\usebox{\plotpoint}}
\multiput(1107,461)(40.946,-6.824){0}{\usebox{\plotpoint}}
\multiput(1113,460)(40.705,-8.141){0}{\usebox{\plotpoint}}
\multiput(1118,459)(40.946,-6.824){0}{\usebox{\plotpoint}}
\multiput(1124,458)(40.705,-8.141){0}{\usebox{\plotpoint}}
\multiput(1129,457)(40.946,-6.824){0}{\usebox{\plotpoint}}
\put(1138.09,455.48){\usebox{\plotpoint}}
\multiput(1141,455)(40.705,-8.141){0}{\usebox{\plotpoint}}
\multiput(1146,454)(40.946,-6.824){0}{\usebox{\plotpoint}}
\multiput(1152,453)(40.705,-8.141){0}{\usebox{\plotpoint}}
\multiput(1157,452)(40.946,-6.824){0}{\usebox{\plotpoint}}
\multiput(1163,451)(40.946,-6.824){0}{\usebox{\plotpoint}}
\multiput(1169,450)(40.705,-8.141){0}{\usebox{\plotpoint}}
\put(1178.95,448.18){\usebox{\plotpoint}}
\multiput(1180,448)(40.705,-8.141){0}{\usebox{\plotpoint}}
\multiput(1185,447)(40.946,-6.824){0}{\usebox{\plotpoint}}
\multiput(1191,446)(40.946,-6.824){0}{\usebox{\plotpoint}}
\multiput(1197,445)(40.705,-8.141){0}{\usebox{\plotpoint}}
\multiput(1202,444)(40.946,-6.824){0}{\usebox{\plotpoint}}
\multiput(1208,443)(40.946,-6.824){0}{\usebox{\plotpoint}}
\multiput(1214,442)(40.705,-8.141){0}{\usebox{\plotpoint}}
\put(1219.81,440.87){\usebox{\plotpoint}}
\multiput(1225,440)(40.705,-8.141){0}{\usebox{\plotpoint}}
\multiput(1230,439)(40.946,-6.824){0}{\usebox{\plotpoint}}
\multiput(1236,438)(40.946,-6.824){0}{\usebox{\plotpoint}}
\multiput(1242,437)(40.705,-8.141){0}{\usebox{\plotpoint}}
\multiput(1247,436)(40.946,-6.824){0}{\usebox{\plotpoint}}
\multiput(1253,435)(40.705,-8.141){0}{\usebox{\plotpoint}}
\put(1260.66,433.56){\usebox{\plotpoint}}
\multiput(1264,433)(40.946,-6.824){0}{\usebox{\plotpoint}}
\multiput(1270,432)(40.705,-8.141){0}{\usebox{\plotpoint}}
\multiput(1275,431)(40.946,-6.824){0}{\usebox{\plotpoint}}
\multiput(1281,430)(40.705,-8.141){0}{\usebox{\plotpoint}}
\multiput(1286,429)(40.946,-6.824){0}{\usebox{\plotpoint}}
\multiput(1292,428)(40.946,-6.824){0}{\usebox{\plotpoint}}
\put(1301.53,426.29){\usebox{\plotpoint}}
\multiput(1303,426)(40.946,-6.824){0}{\usebox{\plotpoint}}
\multiput(1309,425)(41.511,0.000){0}{\usebox{\plotpoint}}
\multiput(1315,425)(40.705,-8.141){0}{\usebox{\plotpoint}}
\multiput(1320,424)(40.946,-6.824){0}{\usebox{\plotpoint}}
\multiput(1326,423)(40.705,-8.141){0}{\usebox{\plotpoint}}
\multiput(1331,422)(40.946,-6.824){0}{\usebox{\plotpoint}}
\put(1342.49,420.09){\usebox{\plotpoint}}
\multiput(1343,420)(40.705,-8.141){0}{\usebox{\plotpoint}}
\multiput(1348,419)(40.946,-6.824){0}{\usebox{\plotpoint}}
\multiput(1354,418)(40.705,-8.141){0}{\usebox{\plotpoint}}
\multiput(1359,417)(40.946,-6.824){0}{\usebox{\plotpoint}}
\multiput(1365,416)(40.946,-6.824){0}{\usebox{\plotpoint}}
\multiput(1371,415)(40.705,-8.141){0}{\usebox{\plotpoint}}
\multiput(1376,414)(40.946,-6.824){0}{\usebox{\plotpoint}}
\put(1383.34,412.73){\usebox{\plotpoint}}
\multiput(1387,412)(40.946,-6.824){0}{\usebox{\plotpoint}}
\multiput(1393,411)(41.511,0.000){0}{\usebox{\plotpoint}}
\multiput(1399,411)(40.705,-8.141){0}{\usebox{\plotpoint}}
\multiput(1404,410)(40.946,-6.824){0}{\usebox{\plotpoint}}
\multiput(1410,409)(40.705,-8.141){0}{\usebox{\plotpoint}}
\multiput(1415,408)(40.946,-6.824){0}{\usebox{\plotpoint}}
\put(1424.29,406.45){\usebox{\plotpoint}}
\multiput(1427,406)(40.705,-8.141){0}{\usebox{\plotpoint}}
\multiput(1432,405)(40.272,-10.068){0}{\usebox{\plotpoint}}
\end{picture}


\caption{Dynamical SGO results for $\zren \phi^{\#}$ as a function of
the inverse heavy-light meson mass $1/M_P^\#$ [27]. 
The upper set of data uses the Clover action for the
light quark (with $\kappa_q = 0.1385$), and the lower set
uses the Wilson action (with $\kappa_q = 0.1585$).
The results of cubic fits in $1/M_P$ to the NRQCD data points
are also shown.}
\end{figure}

\section{Conventional Results}
\label{conv}

The ``conventional'' formalism of the heavy-quark propagator is
simply that obtained from the lattice version of the Dirac
equation, i.e. the same lattice action as for the light
degrees of freedom.
However, potential problems exist when $m_Q a \gtap 1$.
There are two approaches to reduce discretisation effects.
The first is to use an improved action, such as the Clover action \cite{sw},
which has all terms $O(m_Q a)$ explicity removed \cite{grant}.
The second is to use the ``Fermilab'' formulation of
\cite{heavy,paul,andreas,tobe}.
These two choices are not mutually exclusive, and while all
groups use the second technique, one group \cite{jim} uses both.

Table \ref{tab:conv} summarise the data to appear recently
with the references in the second row.
In the following subsections a brief review of each of the four
groups' work is presented.

\begin{table*}[hbt]
\setlength{\tabcolsep}{1.5pc}
\caption{Summary of recent results using the conventional approach.
`Q' signifies Quenched simulation, `D' Dynamical.
See text for detailed comments, and a description of the various
errors.
All numbers should be considered preliminary.}
\label{tab:conv}
\begin{tabular*}{\textwidth}{@{}l@{\extracolsep{\fill}}cccc}
\hline
{\bf Collaboration} & MILC  & \gap LANL - Ohio
                                     & \gap JLQCD & \gap FNAL \\
                    &       & \gap Washington
                                     &        & \\
Reference           &\cite{lat95_claude}
                            & \gap \cite{rajan}
                                     & \gap \cite{lat95_shoji}
                                              & \gap \cite{jim} \\
\hline
\multicolumn{5}{l}{{\bf $\!\!\!\!\!\!\!\!\!\!$Lattice Parameters}}  \\
$n_F$               & 0	\& 2& \gap 0 & \gap 0 & \gap 0 \\
$\beta$             & $5.7 \rightarrow 6.5$ Q
                            & \gap $6.0$
                                     & \gap $6.1,6.3$
                                              & \gap $5.9$ \\
                    & $5.4 \rightarrow 5.7$ D &&& \\
Volume              & $\le 32^3 \times 100$
                            & \gap $32^3 \times 64$
                                     & \gap $\le 32^3 \times 80$
                                              & \gap $16^3 \times 32$ \\
$N_{cfgs}$          &$\sim100$
                            & \gap $170$
                                     & \gap $\sim100$
                                              & \gap $100$ \\
$S_q$               & Wilson& \gap Wilson & \gap Wilson & \gap Clover \\
$S_Q$               & Fermilab-Wilson
                            & \gap Fermilab-Wilson
                                     & \gap Fermilab-Wilson
                                              & \gap Fermilab-Clover \\
                    & \& Static&     &        & \\
\hline
{\bf Results}       &       &        &        & \\
$f_D$ [MeV]         & $182\pm3(9)(22)$
                            & \gap $229\pm7(-)(+12)(^{+18}_{-14})(7)$
                                     & \gap $214\pm10(25)$
                                              & \gap $220^{+4}_{-5}$ \\
$f_{D_s}$           & $198\pm5(10)(19)$
                            & \gap $260\pm4(-5)(+15)(^{+24}_{-20})(8)$
                                     & \gap $248\pm7(29)$
                                              & \gap $(239^{+3}_{-4})$ \\
$\fb$ [MeV]         & $151\pm5(16)(26)$
                            &        & \gap $221\pm15(26)$
                                              & \gap $188^{+6}_{-4})$ \\
$f_{B_S}$           & $169\pm7(14)(29)$
                            &        & \gap $244\pm8(28)$
                                              & \gap $(207^{+3}_{-2})$ \\
\hline
\end{tabular*}
\end{table*}

\subsection{MILC Results}
\label{milc}

The MILC collaboration's results \cite{lat95_claude} distinguish themselves
because they are the only simulations in the conventional
approach that (i) have dynamical fermion runs,
(ii) have a static point to help constrain their fits in $1/M_P$,
and (iii) use the hopping parameter expansion to give them
a wide range of $\kappa_Q$ values at very little cost \cite{david}.
They have also studied finite volume effects, and performed
a continuum extrapolation. (This last point will be discussed
more later.)

In fig.6 
the quantity $\phi$ is plotted
against $1/M_P$, both in physical units.
The curve shown is a fit to the points marked with a cross
(i.e. in this case, points with $M_P < M_D$). By varying the fitting window
in $M_P$, an estimate is obtained of the error due to the $m_Q a$ effects
which escape the Fermilab redefinitions.
Another estimate of this error would be to fit the standard
Wilson data (without the Fermilab redefinitions) for, say,
$M_P < M_D$, and compare it with the results from Fermilab data
for the same $M_P$ range.

Fig.7 
shows the continuum
extrapolation of $\fb$ with $f_{\pi}$ used to set the scale.
Both quenched and dynamical data are shown.
The line shown is a linear fit to all the quenched data
(i.e. $5.7 \le \beta \le 6.5$).
It is probable that this procedure suffers from the same
problem discussed in sec. \ref{cont_limit}, where static
data for $\beta \ltap 6.0$ was found not to scale, within
statistical errors.
If this scenario is correct, then including the data with
$\beta \ltap 6.0$ in a linear fit skews the continuum extrapolation
downwards.
Fig.7 
seems to confirm this assertion, though the effect is small.
This would explain at least part of the difference between the
MILC results and the other data shown in table \ref{tab:conv}.

The dynamical data points in the figure show a clear discrepancy
at around $a=0.5 \mbox{GeV}^{-1}$ compared with the quenched data.
This can be used as an estimate of the quenching errors which
appears to be considerable at this stage \cite{lat95_claude}.

The results of the MILC collaboration's work is displayed in
table \ref{tab:conv}. The three errors stated are:
(i) statistical, (ii) various systematic, and (iii) quenching.
The systematic errors are a combination of the uncertainties
in the various fits, the $m_Q a$ effects, finite volume and
finite $a$ errors.

\begin{figure}[t]
\vspace{-35mm}
\protect\label{fig:phi_milc}.

\psfig{file=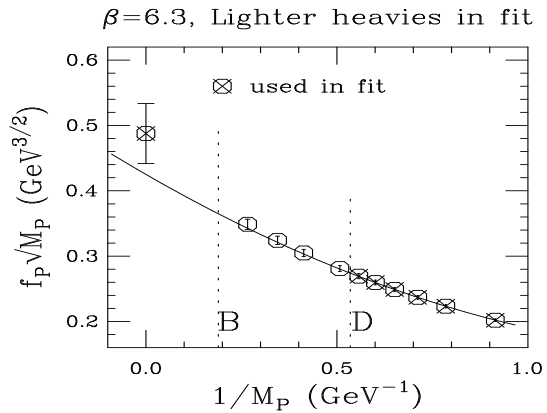,width=80mm,height=90mm,angle=0,clip=}

\vspace{-10mm}
\caption{MILC results for $\phi$ as a function of
$1/M_P$ [28]. 
The data points used in this particular fit are
denoted with a cross (see text).}
\vspace{-5mm}
\end{figure}


\begin{figure}
\vspace{-35mm}
\protect\label{fig:fb_milc}.

\psfig{file=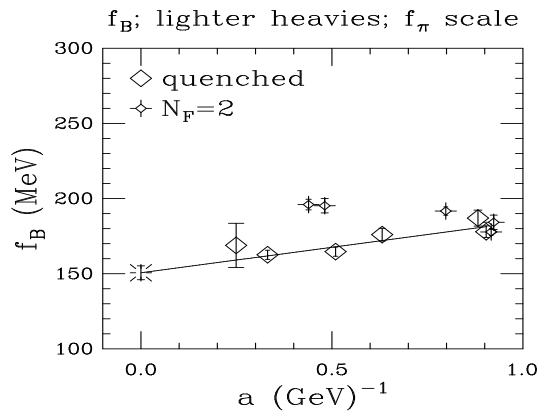,width=80mm,height=90mm,angle=0,clip=}

\vspace{-10mm}
\caption{MILC results for $\fb$ as a function of
lattice spacing $a$ [28]. 
The fit is to the quenched points only (see text).}
\vspace{-5mm}
\end{figure}


\subsection{LANL-Ohio-Washington Results}
\label{lanl}

The LANL-Ohio-Washington collaboration have new results using the
Fermilab - Wilson formulation \cite{rajan}. Since they have no static
point, and are working at a moderate $\beta$ value, no estimate of
$\fb$ was given.
Of particular interest is their
comprehensive study of the systematics involved in
setting the strange and charm quark masses, and in
the definition of $\zren$. This is outlined below.

\begin{itemize}
\item
The hopping parameter corresponding to the strange
quark mass, $\kappa_s$, is defined in three different ways: by fixing
$M_K^2/M_\pi^2$; $M_K^*/M_\rho$; and $M_\phi/M_\rho$
to be equal to their physical values.
These estimates correspond to a surprisingly large $\sim 20\%$
spread in the values of the strange quark mass.

\item
The charm quark is set by fixing the mass of a meson
containing a c-quark to its experimental value.
The lattice estimate of these heavy-light meson masses
is defined using either the pole mass (obtained
from the exponential decay of the correlation functions),
or by the lattice dispersion relation.

\item
A careful study of the systematic effects entering the definition
of $\zren$ was undertaken. Here different values of the
Lepage-Mackenzie $q^*$ were used \cite{lm}, and two
definitions ($u_0 = 1/8\kappa_c$ and $<U_{plaq}>^{1/4}$) were
used to redefine the coupling.
\end{itemize}

The spread of decay constant values using each of these methods
gives estimates of the corresponding systematic effects.
The final results for $f_D$ and $f_{D_s}$ are shown in
table \ref{tab:conv}. The five errors listed are due to:
(i) statistics, (ii) strange quark determination,
(iii) charm quark determination, (iv) $\zren$, (v) setting the scale.
The importance of this work is that it shows that each of the
three errors outlined above are at least of the same order
as the statistical and scale errors (which are normally assumed
to be the dominant errors).

\subsection{JLQCD Results}

A status report of the ongoing analysis by the JLQCD collaboration
was presented at this conference \cite{lat95_shoji}.
They have data at two $\beta$ values and use
a quadratic fit in $1/M_P$ to determine their values for $f_B$, $f_D$ etc.
Their results, taken from the $\beta=6.3$ dataset are
given in table \ref{tab:conv}.
The $\rho$ mass was used to set the scale.
The first quoted error is statistical, and the second is
due to the scale. More accurate results will be obtained from
this collaboration in the near future.

\subsection{FNAL Results}

The FNAL group are simulating the Clover action in the Fermilab
formalism at $\beta=5.9$ \cite{jim}. They have preliminary results
which are shown in table \ref{tab:conv}. (Statistical errors only
are shown.) The strong point of this work
is that it uses the Clover action (i.e. complete with the rotations of the
quark fields) and that therefore the results should be free of $O(m_Q
a)$ effects. Values for the decay constants in the table use $f_\pi$ to
set the scale. $f_{D_s}$ and ${f_{B_S}}$ are quoted in parentheses,
since the strange quark value used in their determination is nominal.
The FNAL group plan to simulate at other $\beta$ values to check
the scaling of these quantities.


\section{Conclusions}

One of the main themes of this review is that
systematic effects now clearly dominate statistical
errors in lattice calculations of $f_P$.
In many ways this is a very desirable situation
since it allows their careful callibration.
I list the main systematic uncertainties and their possible
solutions.

\begin{itemize}
\item
$\zren$ varies by as much as $\sim 30\%$ (see secs. \ref{zren_stat}
and \ref{lanl}).
This variation can be significantly improved by using the clover action.
The non-perturbative value of $\zren$
will clear up this uncertainty \cite{lat95_mauro}.

\item
The continuum limit of $f_P$ can be studied
by an $a \rightarrow 0$ extrapolation.
In the static case, only data with $\beta \ltap 6.0$
should be included in this fit (for present levels of statistics,
see sec. \ref{cont_limit}), otherwise the $a=0$ value will be
biased downwards.
Presumably a remnant of this effect survives in the
NRQCD and conventional cases.
Clearly $O(a)$ effects should be reduced by the use of the clover action.

\item
Dynamical ($n_F=2$) results seem to produce higher values for $f_P$
for both the NRQCD (sec. \ref{sgo_dyn}) and conventional cases
(sec. \ref{milc}).
Whether this effect is invariant under further investigation
will be a main source of research over the next couple of years.
A more subtle question is the extent to which $n_F=2$ data
truly reflects the real $n_F \approx 3$ world \cite{steve,booth}.

\item
The choice of physical quantity to set $a$ impacts upon the final
value of $f_P$. Again, short-term, this should be fed into the
systematic uncertainties, but longer-term this should be
understood in terms of quenching and/or $O(a)$ errors.

\item
The chiral extrapolation of methods which use a {\em fixed} smearing
for all light quark values should be questioned (see \ref{ape}).
This is not a large effect for $f_P$ but can become significant for
chiral ratios/differences such as $f_{B_S}/f_B$ and $M_{B_S}-M_B$.

\item
Setting the hopping parameters corresponding to the
physical strange and charm quarks also introduces
an error which should be included (see sec. \ref{lanl}).

\end{itemize}

The wide range of $M_P$ values presently covered by the
three methods allow an important consistency test to be
made: Do the different approaches agree with eachother?
Studying the $f_B$ values in tables \ref{tab:stat},\ref{tab:nrqcd}
and \ref{tab:conv}, together with the $\phi$ versus $1/M_P$
plots suggests that they do.
Note that there is still a fair degree of `slop' in the data
due to the fact that different groups have used different
definitions of $\zren$ and chosen different quantities to
set the scale. Once this is tightened up, this consistency
test can be made more definitive (e.g. by using
{\em all three} methods to determine $f_P$ on the same
configurations).

Using the data in the tables I estimate the following ``global''
lattice averages: $f_D \approx f_B = 200 \mbox{MeV} \pm 20\%$.
The error bar includes my estimate of all uncertainties.

\section{Acknowledgements}

I wish to thank my collaborators M. Crisafulli, V. Lubicz, G. Martinelli,
F. Rapuano and A. Vladikas for many useful discussions over the last
several stimulating years in Rome.
Supported from the EU Human Capital and Mobility
Programme, Grant ERBCHBICT 941462 is much appreciated.



\end{document}